\documentclass[journal,draftcls,onecolumn,letterpaper,twoside,11pt]{IEEEtran}

\usepackage[cmex10]{amsmath}
\usepackage{latexsym,amssymb,cite,bbm,epsf,amsthm,subfigure,algorithm,algorithmic}
\usepackage{graphicx}
\usepackage{enumerate}
\usepackage{mathrsfs}
\usepackage{epstopdf}
\usepackage{bbm}
\usepackage{bigints}

\newcommand{\bc}{\begin{center}}
\newcommand{\ec}{\end{center}}
\newcommand{\be}{\begin{equation}}
\newcommand{\ee}{\end{equation}}
\newcommand{\ba}{\begin{array}}
\newcommand{\ea}{\end{array}}
\newcommand{\bea}{\begin{eqnarray}}
\newcommand{\eea}{\end{eqnarray}}
\newcommand{\bal}{\begin{align}}
\newcommand{\eal}{\end{align}}
\newcommand{\ei}{\end{itemize}}
\newcommand{\bi}{\begin{itemize}}
\newcommand{\bfi}{\begin{figure}}
\newcommand{\efi}{\end{figure}}
\newcommand{\MB}{\left[\begin{array}}
\newcommand{\ME}{\end{array}\right]}
\newcommand{\nn}{\nonumber}

\newtheorem{thm}{Theorem}

\newtheorem{cor}{Corollary}
\newtheorem{lem}{Lemma}
\newtheorem{pro}{Proposition}

\renewcommand{\vec}[1]{\mbox{\boldmath${#1}$}}

\newcommand{\Exp}{\mathsf{E}}
\newcommand{\Var}{\mathsf{Var}}
\newcommand{\bExp}{\bar{\mathsf{E}}}
\newcommand{\Pro}{\mathsf{P}}
\newcommand{\Hyp}{\mathsf{H}}

\newcommand{\x}{\vec{\mathsf{x}}}
\newcommand{\T}{\mathsf{T}}
\newcommand{\sd}{\mathsf{d}}
\newcommand{\Tr}{\mathsf{Tr}}
\newcommand{\Cov}{\mathsf{Cov}}

\newcommand{\cR}{\mathcal{R}}

\newcommand{\cH}{\mathcal{H}}
\newcommand{\cF}{\mathcal{F}}

\newcommand{\cN}{\mathcal{N}}

\newcommand{\cC}{\mathcal{C}}

\newcommand{\sC}{\mathscr{C}}

\newcommand{\bN}{\mathbb{N}}
\newcommand{\bR}{\mathbb{R}}

\newcommand{\vone}{\vec{1}}

\newcommand{\vy}{\vec{y}}
\newcommand{\vx}{\vec{x}}
\newcommand{\vw}{\vec{w}}

\newcommand{\ve}{\vec{e}}
\newcommand{\vv}{\vec{v}}
\newcommand{\vmu}{\vec{\mu}}

\newcommand{\mH}{\vec{H}}
\newcommand{\mI}{\vec{I}}

\newcommand{\mU}{\vec{U}}
\newcommand{\mSi}{\vec{\Sigma}}

\newcommand{\ind}[1]{\mathbbm{1}_{\{#1\}}}   %indicator

\newcommand{\ignore}[1]{{}}

\usepackage{tikz}

\begin{document}

\title{Sequential Joint Detection and Estimation: Optimum Tests and Applications}

\author{Yasin~Y{\i}lmaz, %~\IEEEmembership{Student Member,~IEEE,}
Shang~Li, %~\IEEEmembership{}
and~Xiaodong~Wang%,~\IEEEmembership{Fellow,~IEEE}
%\thanks{Manuscript received December 24, 2013; revised May 8, 2014. This work was supported in part by the U.S. National Science Foundation under grant CIF1064575, and in part by the U.S. Office of Naval Research under grant N000141210043.
\thanks{The authors are with the Department of Electrical Engineering, Columbia University, New York, NY 10027 USA (e-mail: \{yasin,shang,wangx\}@ee.columbia.edu).}}
%X. Wang is with the Electrical Engineering Department, Columbia University, New York, NY 10027 USA, and also with King Abdulaziz University, Jeddah, Saudi Arabia (e-mail: wangx@ee.columbia.edu).}}

\maketitle

\begin{abstract}
We treat the statistical inference problems in which one needs to detect and estimate simultaneously using as small number of samples as possible. Conventional methods treat the detection and estimation subproblems separately, ignoring the intrinsic coupling between them. However, a joint detection and estimation problem should be solved to maximize the overall performance. We address the sample size concern through a sequential and Bayesian setup. Specifically, we seek the optimum triplet of stopping time, detector, and estimator(s) that minimizes the number of samples subject to a constraint on the combined detection and estimation cost. A general framework for optimum sequential joint detection and estimation is developed. The resulting optimum detector and estimator(s) are strongly coupled with each other, proving that the separate treatment is strictly sub-optimum. The theoretical results derived for a quite general model are then applied to several problems with linear quadratic Gaussian (LQG) models, including dynamic spectrum access in cognitive radio, and state estimation in smart grid with topological uncertainty. Numerical results corroborate the superior overall detection and estimation performance of the proposed schemes over the conventional methods that handle the subproblems separately.
\end{abstract}


\section{Introduction}
\label{sec:intro}

Detection and estimation problems appear simultaneously in a wide range of fields, such as wireless communications, power systems, image processing, genetics, and finance. For instance, to achieve effective and reliable dynamic spectrum access in a cognitive radio system, a secondary user needs to detect primary user transmissions, and if detected to estimate the cross channels that may cause interference to primary users \cite{Yilmaz_dsa}. In power grid monitoring, it is essential to detect the correct topological model, and at the same time estimate the system state \cite{Chen13}. Some other important examples are detecting and estimating objects from images \cite{Vo10}, target detection and parameter estimation in radar \cite{Tajer10}, and detection and estimation of periodicities in DNA sequences \cite{Jaja12}.

In all these applications, detection and estimation problems are intrinsically coupled, and are both of primary importance. Hence, a jointly optimum method, that maximizes the overall performance, is needed.
Classical approaches either treat the two subproblems separately with the corresponding optimum solutions, or solve them together, as a composite hypothesis testing problem, using the generalized likelihood ratio test (GLRT).
However, such approaches do not yield the overall optimum solution \cite{Middleton68,Yilmaz_jde}.
In the former approach, for example, the likelihood ratio test (LRT) is performed by averaging over the unknown parameters to solve the detection subproblem optimally; and then based on the detection decision, the Bayesian estimators are used to solve the estimation subproblem.
On the other hand, in GLRT, the maximum likelihood (ML) estimates of all unknown parameters are computed, and then using these estimates, the LRT is performed as in a simple hypothesis testing problem. In GLRT, the primary emphasis is
on the detection performance and the estimation performance is of secondary importance. GLRT is very popular due to its simplicity. However, even its detection performance is not optimal in the Neyman-Pearson sense \cite{Zeitouni92}, and neither is the overall performance under mixed Bayesian/Neyman-Pearson \cite{Moustaki11} and pure Bayesian \cite{Middleton68} setups.

The first systematic theory on joint detection and estimation appeared in \cite{Middleton68}. This initial work, in a Bayesian framework, derives optimum joint detector and estimator structures for different levels of coupling between the two subproblems. \cite{Fredriksen72} extends the results of \cite{Middleton68} on binary hypothesis testing to the multi-hypothesis case. In \cite{Birdsall73}, different from \cite{Middleton68,Fredriksen72}, the case with unknown parameters under the null hypothesis is considered. \cite{Birdsall73} does not present an optimum joint detector and estimator, but shows that, even in the classical separate treatment of the two subproblems, LRT implicitly uses the posterior distributions of unknown parameters, which characterize the Bayesian estimation. \cite{Baygun95} deals with joint multi-hypothesis testing and non-Bayesian estimation considering a finite discrete parameter set and the minimax approach. \cite{Moustaki11} and \cite{Moustaki12} study Bayesian estimation under different Neyman-Pearson-like formulations, and derive the corresponding optimum joint detection and estimation schemes. \cite{Jaja12}, in a minimax sense, extends the analysis in \cite{Moustaki12} to the general case with unknown parameters in both hypotheses. \cite{Chen13} handles the joint multi-hypothesis testing and state estimation problem for linear models with Gaussian noise. It finds the joint posterior distribution of the hypotheses and the system states, which can be used to identify the optimum joint detector and estimator for a specific performance criterion in a unified Bayesian approach.

Most of the today's engineering applications are subject to resource (e.g., time, energy, bandwidth) constraints. For that reason, it is essential to minimize the number of observations used to perform a task (e.g., detection, estimation) due to the cost of taking a new observation, and also latency constraints. Sequential statistical methods are designed to minimize the average number of observations for a given accuracy level. They are equipped with a stopping rule to achieve optimal stopping, unlike fixed-sample-size methods. Specifically, we cannot stop taking samples too early due to the performance constraints, and do not want to stop too late to save critical resources, such as time and energy. Optimal stopping theory handles this trade-off through sequential methods. For more information on sequential methods we refer to the original work \cite{Wald47} by Wald, and a more recent book \cite{Govin04}. The majority of existing works on joint detection and estimation consider only the fixed-sample-size problem. Although \cite{Birdsall73} discusses the case where observations are taken sequentially, it does not consider optimal stopping, limiting the scope of the work to the iterative computation of sufficient statistics. The only work that treats the joint detection and estimation problem in a ``real" sequential manner is \cite{Yilmaz_jde}. It provides the exact optimum triplet of stopping time, detector, and estimator for a linear scalar observation model with Gaussian noise, where there is an unknown parameter only under the alternative hypothesis.

In this paper, we solve the optimum sequential joint detection and estimation problem under the most general setup, namely for a general non-linear vector signal model with arbitrary noise distribution and unknown parameters under both hypotheses. We also do not assume a specific estimation cost function.

The remainder of the paper is organized as follows. In Section \ref{sec:sjde}, we derive the optimum procedure for sequential joint detection and estimation under a general setup. We then apply the theory developed in Section \ref{sec:sjde} to a general linear quadratic Gaussian model in Section \ref{sec:lin}, dynamic spectrum access in cognitive radio networks in Section \ref{sec:cr}, and state estimation in smart grid with topological uncertainty in Section \ref{sec:sg}. Finally, concluding remarks are given in Section \ref{sec:conc}.

\section{Optimum Sequential Joint Detection and Estimation}
\label{sec:sjde}

\subsection{Problem Formulation}

Consider a general model
\be
\label{eq:sig_nonlin}
\vy_t = f(\vx,\mH_t) + \vw_t, ~t=1,2,\ldots,
\ee
where $\vy_t\in\bR^M$ is the measurement vector taken at time $t$; $\vx\in\bR^N$ is the unknown vector of parameters that we want to estimate; $\mH_t$ %\in\bR^{m\times n}$
is the observation matrix that relates $\vx$ to $\vy_t$; $f$ is a (possibly nonlinear) function of $\vx$ and $\mH_t$; and $\vw_t\in\bR^M$ is the noise vector.

In addition to estimation, we would like to detect the true hypothesis ($\Hyp_0$ or $\Hyp_1$) in a binary hypothesis testing setup, in which $\vx$ is distributed according to a specific probability distribution under each hypothesis, i.e.,
\begin{align}
\label{eq:hypo}
\begin{split}
	\Hyp_0&:  \vx\sim \pi_0, \\
	\Hyp_1&:  \vx\sim \pi_1.
\end{split}
\end{align}
Here, we do not assume specific probability distributions for $\vx$, $\mH_t$, $\vw_t$, or a specific system model $f$. Moreover, we allow for correlated noise $\vw_t$ and correlated $\mH_t$. We only assume $\pi_0$ and $\pi_1$ are known, and $\{\vy_t,\mH_t\}$ are observed at each time $t$. Note that random and observed $\mH_t$ is a more general model than deterministic and known $\mH_t$. 
%This is a more general setup than the one in \cite{Yilmaz_jde}, in which $x=0$, i.e., $\mu_0=\sigma_0^2=0$.
We denote with $\cH_t$ and $\{\cH_t\}$ the sigma-algebra and filtration generated by the history of the observation matrices $\{\mH_1,\ldots,\mH_t\}$, respectively, and with $\Pro_i$ and $\Exp_i$ the probability measure and expectation under $\Hyp_i$. 

Since we want to both detect and estimate, we use a combined cost function
\begin{multline}
\label{eq:cost}
	\sC(T,d_{T},\hat{\vx}_{T}^0,\hat{\vx}_{T}^1) = a_0 \Pro_0(d_{T}=1|\cH_{T}) + a_1 \Pro_1(d_{T}=0|\cH_{T}) \\
+ b_{00} \Exp_0\left[J(\hat{\vx}_{T}^0,\vx) \ind{d_{T}=0}|\cH_{T}\right] + b_{01} \Exp_0\left[J(\hat{\vx}_{T}^1,\vx) \ind{d_{T}=1}|\cH_{T}\right] \\
+ b_{10} \Exp_1\left[J(\hat{\vx}_{T}^0,\vx) \ind{d_{T}=0}|\cH_{T}\right] + b_{11} \Exp_1\left[J(\hat{\vx}_{T}^1,\vx) \ind{d_{T}=1} |\cH_{T}\right]
\end{multline}
where $T$ is the stopping time, $d_{T}$ is the detection function, $\{\hat{\vx}_{T}^0,\hat{\vx}_{T}^1\}$ are the estimators when we decide on $\Hyp_0$ and $\Hyp_1$, respectively, $J(\hat{\vx}_{T},\vx)$ is a general estimation cost function, e.g., $\|\hat{\vx}_{T}-\vx\|^2$, and $\{a_i,b_{ij}\}_{i,j=0,1}$ are some constants. The indicator function $\ind{A}$ takes the value $1$ if the event $A$ is true, or $0$ otherwise. In \eqref{eq:cost}, the first two terms are the detection cost, and the remaining ones are the estimation cost. Writing \eqref{eq:cost} in the following alternative form
\begin{multline}
\label{eq:cost1}
	\sC(T,d_{T},\hat{\vx}_{T}^0,\hat{\vx}_{T}^1) = \Exp_0\left[b_{00}J(\hat{\vx}_{T}^0,\vx) \ind{d_{T}=0} + \left\{a_0+b_{01}J(\hat{\vx}_{T}^1,\vx)\right\} \ind{d_{T}=1}|\cH_{T}\right] \\
+ \Exp_1\left[\left\{a_1+b_{10}J(\hat{\vx}_{T}^0,\vx)\right\} \ind{d_{T}=0} + b_{11}J(\hat{\vx}_{T}^1,\vx) \ind{d_{T}=1} |\cH_{T}\right] %\\
%	= p_0 ~\Exp_0\left[\tilde{b}_{00}J(\hat{\vx}_{T}^0,\vx) \ind{d_{T}=0} + \left\{\tilde{a}_0+\tilde{b}_{01}J(\hat{\vx}_{T}^1,\vx)\right\} \ind{d_{T}=1}|\cH_{T}\right] \\
%+ p_1 ~\Exp_1\left[\left\{\tilde{a}_1+\tilde{b}_{10}J(\hat{\vx}_{T}^0,\vx)\right\} \ind{d_{T}=0} + \tilde{b}_{11}J(\hat{\vx}_{T}^1,\vx) \ind{d_{T}=1} |\cH_{T}\right]
\end{multline}
it is clear that our cost function corresponds to the Bayes risk given $\{\mH_1,\ldots,\mH_t\}$.

In a sequential setup, in general, the expected stopping time (i.e., the average number of samples) is minimized subject to a constraint on the cost function. In the presence of an auxiliary statistic, such as $\cH_t$, conditioning is known to have significant advantages \cite{Efron78}, hence the cost function in \eqref{eq:cost} is conditioned on $\cH_t$. Intuitively, there is no need to average the performance measure $\sC(T,d_{T},\hat{\vx}_{T}^0,\hat{\vx}_{T}^1)$ over $\cH_t$, which is an observed statistic. Conditioning on $\cH_t$ also frees our formulation from assuming statistical descriptions (e.g., probability distribution, independence, stationarity) on the observation matrices $\{\mH_t\}$.
As a result, our objective is to minimize $\Exp[T|\cH_t]$ subject to a constraint on $\sC(T,d_{T},\hat{\vx}_{T}^0,\hat{\vx}_{T}^1)$.

Let $\cF_t$ and $\{\cF_t\}$ denote the sigma-algebra and filtration generated by the complete history of observations $\{(\vy_1,\mH_1),\ldots,(\vy_t,\mH_t)\}$, respectively, thus $\cH_t\subset\cF_t$. In the pure detection and pure estimation problems, it is well known that serious analytical complications arise if we consider a general $\{\cF_t\}$-adapted stopping time, that depends on the complete history of observations. Specifically, in the pure estimation problem, finding the optimum sequential estimator that attains the sequential Cramer-Rao lower bound (CRLB) is not a tractable problem if $T$ is adapted to the complete observation history $\{\cF_t\}$ \cite{Ghosh87,Ghosh91}. Similarly, in the pure detection problem with an $\{\cF_t\}$-adapted stopping time, we end up with a two-dimensional optimal stopping problem which is impossible to solve (analytically) since the thresholds for the running likelihood ratio depend on the sequence $\{\mH_t\}$. Alternatively, in \cite{Grambsch83,Fellouris12,Yilmaz_est,Yilmaz_jde}, $T$ is restricted to $\{\cH_t\}$-adapted stopping times, which facilitates obtaining an optimal solution.
In this paper, we are interested in $\{\cH_t\}$-adapted stopping times as well. %Furthermore, we formulate the problem conditioned on the ancillary statistic $\cH_t$ [cf. \eqref{eq:cost}], whose significant advantage over the unconditional formulation was shown in \cite{Efron78} and more recently in \cite{Yilmaz_vest}.
Hence, $\Exp[T|\cH_t]=T$ and we aim to solve the following optimization problem,
\be
\label{eq:pro}
	\min_{T,d_{T},\hat{\vx}_{T}^0,\hat{\vx}_{T}^1} T ~~~~~ \text{subject to} ~~~ \sC(T,d_{T},\hat{\vx}_{T}^0,\hat{\vx}_{T}^1) \leq \alpha,
\ee
where $\alpha$ is a target accuracy level.

From an operational point of view, we start with the following stopping rule: stop the first time the target accuracy level $\alpha$ is achieved, i.e., the inequality $\sC(T,d_{T},\hat{\vx}_{T}^0,\hat{\vx}_{T}^1) \leq \alpha$ is satisfied. This operational problem statement gives us the problem formulation in \eqref{eq:pro}, which in turn defines an $\{\cH_t\}$-adapted stopping time $T$. This is because $T$ is solely determined by $\sC(T,d_{T},\hat{\vx}_{T}^0,\hat{\vx}_{T}^1)$, which, as seen in \eqref{eq:cost}, averages over $\{\vy_t\}$ and thus is a function of only $\{\mH_t\}$. The stopping rule considered here is a natural extension of the one commonly used in sequential estimation problems, e.g., \cite{Grambsch83,Yilmaz_est}, and is optimum for $\{\cH_t\}$-adapted stopping times, as shown in \eqref{eq:pro}. Note that the solution sought in \eqref{eq:pro} is optimum for each realization of $\{\mH_t\}$, and not on average with respect to this sequence.

\subsection{Optimum Solution}
\label{sec:soln}

\subsubsection*{\underline{Optimum Estimators}}
Let us begin our analysis with the optimum estimators. 

\begin{lem}
\label{lem:est}
The optimum estimators $\hat{\x}_T^0$ and $\hat{\x}_T^1$ for the problem in \eqref{eq:pro} are given by
\be
\label{eq:est}
    \hat{\x}_T^i = \arg\min_{\hat{\vx}} \bExp_i\left[J(\hat{\vx},\vx) |\cF_T\right],~i=0,1,
\ee
where $\bExp_i$ is the expectation under the probability distribution 
\be
\label{eq:pro_mix}
\bar{p}_t^i(\vx|\cF_t) \triangleq \frac{b_{0i}p_0(\vx|\cF_t) + b_{1i}L_t p_1(\vx|\cF_t)}{b_{0i}+b_{1i}L_t},
\ee
$p_i(\vx|\cF_t)$ is the posterior distribution under $\Hyp_i$, and 
\be
\label{eq:like}
L_t \triangleq \frac{p_1(\{\vy_s\}_{s=1}^t|\cH_t)}{p_0(\{\vy_s\}_{s=1}^t|\cH_t)}
\ee
is a likelihood ratio. Specifically, the minimum mean-squared error (MMSE) estimator, for which \\ $J(\hat{\vx},\vx)=\|\hat{\vx}-\vx\|^2$, is given by
\be
\label{eq:est_mmse}
    \hat{\x}_T^i = \frac{b_{0i}\Exp_0[\vx|\cF_T] + b_{1i}L_T \Exp_1[\vx|\cF_T]}{b_{0i}+b_{1i}L_T},~i=0,1.
\ee
\end{lem}

\begin{IEEEproof}
If we find a pair of estimators that minimize the cost function $\sC(T,d_{T},\hat{\vx}_{T}^0,\hat{\vx}_{T}^1)$ for any stopping time $T$ and detector $d_{T}$, then, from \eqref{eq:pro}, these estimators are the optimum estimators $(\hat{\x}_T^0,\hat{\x}_T^1)$.
Grouping the terms with the same estimator in \eqref{eq:cost}, we can write the optimum estimators as
\begin{align}
    \hat{\x}_T^0 &= \arg\min_{\hat{\vx}} b_{00} \Exp_0\left[J(\hat{\vx},\vx) \ind{d_{T}=0}|\cH_{T}\right] + b_{10} \Exp_1\left[J(\hat{\vx},\vx) \ind{d_{T}=0} |\cH_{T}\right] \nn\\
    \hat{\x}_T^1 &= \arg\min_{\hat{\vx}} b_{01} \Exp_0\left[J(\hat{\vx},\vx) \ind{d_{T}=1}|\cH_{T}\right] + b_{11} \Exp_1\left[J(\hat{\vx},\vx) \ind{d_{T}=1} |\cH_{T}\right]. \nn
\end{align}
Using the likelihood ratio 
$$
\bar{L}_T \triangleq \frac{p_1(\{\vy_s\}_{t=1}^T,\vx|\cH_T)}{p_0(\{\vy_s\}_{t=1}^T,\vx|\cH_T)}
$$ 
we can write 
\be
\Exp_1\left[J(\hat{\vx},\vx) \ind{d_{T}=0} |\cH_{T}\right] = \Exp_0\left[\bar{L}_T J(\hat{\vx},\vx) \ind{d_{T}=0} |\cH_{T}\right], \nn
\ee
and accordingly
\be
\hat{\x}_T^0 = \arg\min_{\hat{\vx}} \Exp_0\Big[ \big( b_{00} + b_{10} ~\bar{L}_T \big) J(\hat{\vx},\vx) \ind{d_{T}=0} |\cH_T \Big]. \nn
\ee

To free the expectation from random $T$ we first rewrite the above equation as
\be
\hat{\x}_T^0 = \arg\min_{\hat{\vx}} \Exp_0\Big[ \sum_{t=0}^{\infty} \big( b_{00} + b_{10} ~\bar{L}_t \big) J(\hat{\vx},\vx) \ind{d_{t}=0} \ind{T=t} |\cH_{t} \Big], \nn
\ee
then take $\ind{T=t}$ outside the expectation 
\be
\hat{\x}_T^0 = \arg\min_{\hat{\vx}} \sum_{t=0}^{\infty} \Exp_0\Big[ \big( b_{00} + b_{10} ~\bar{L}_t \big) J(\hat{\vx},\vx) \ind{d_{t}=0} |\cH_{t} \Big] \ind{T=t}, \nn
\ee
as $T$ is $\{\cH_t\}$-adapted, hence $\ind{T=t}$ is $\cH_t$-measurable, i.e., deterministic given $\cH_t$.

Recall that $\cF_t$ denotes the sigma-algebra generated by the complete history of observations \\$\{(\vy_1,\mH_1),\ldots,(\vy_t,\mH_t)\}$, and thus $\cH_t\subset\cF_t$. Since $\Exp_0[~\cdot~|\cH_t]=\Exp_0\Big[~\Exp_0[~\cdot~|\cF_t]~\big|\cH_t\Big]$, we write
\be
\hat{\x}_T^0  = \arg\min_{\hat{\vx}} \sum_{t=0}^{\infty} \Exp_0\Big[ b_{00} \Exp_0\left[J(\hat{\vx},\vx) \ind{d_{t}=0} |\cF_{t}\right] + b_{10} \Exp_0\left[\bar{L}_t~J(\hat{\vx},\vx) \ind{d_{t}=0} |\cF_{t}\right] \big|\cH_{t} \Big] \ind{T=t}.\nn
\ee
Note that $d_t$ is $\cF_t$-measurable, i.e., a feasible detector is a function of the observations only, hence deterministic given $\cF_t$. Then, we have 
\be
\hat{\x}_T^0  = \arg\min_{\hat{\vx}} \sum_{t=0}^{\infty} \Exp_0\Big[ \big\{ b_{00} \Exp_0\left[J(\hat{\vx},\vx) |\cF_{t}\right] + b_{10}\Exp_0\left[\bar{L}_t ~J(\hat{\vx},\vx) |\cF_{t}\right] \big\} \ind{d_{t}=0} \big|\cH_{t} \Big] \ind{T=t}, \nn
\ee
which reduces to
\be
\label{eq:est_sep}
\hat{\x}_T^0  = \arg\min_{\hat{\vx}} \sum_{t=0}^{\infty} \left\{ b_{00} \Exp_0\left[J(\hat{\vx},\vx) |\cF_{t}\right] + b_{10}\Exp_0\left[\bar{L}_t ~J(\hat{\vx},\vx) |\cF_{t}\right] \right\} \ind{T=t}.
\ee

Expand the likelihood ratio $\bar{L}_t$ as
\begin{align}
\bar{L}_t &= \frac{p_1(\{\vy_s\}_{s=1}^t,\vx|\cH_t)}{p_0(\{\vy_s\}_{s=1}^t,\vx|\cH_t)} \nn\\
&= \frac{p_1(\{\vy_s\}_{s=1}^t|\cH_t)}{p_0(\{\vy_s\}_{s=1}^t|\cH_t)}~ \frac{p_1(\vx|\{\vy_s\}_{s=1}^t,\cH_t)}{p_0(\vx|\{\vy_s\}_{s=1}^t,\cH_t)} \nn\\
&= \frac{p_1(\{\vy_s\}_{s=1}^t|\cH_t)}{p_0(\{\vy_s\}_{s=1}^t|\cH_t)}~ \frac{p_1(\vx|\cF_t)}{p_0(\vx|\cF_t)}, \nn
\end{align}
and denote the first term above with
\be
L_t = \frac{p_1(\{\vy_s\}_{s=1}^t|\cH_t)}{p_0(\{\vy_s\}_{s=1}^t|\cH_t)}, \nn
\ee 
which is also a likelihood ratio. Given $\cF_t$, $L_t$ is deterministic, hence in \eqref{eq:est_sep}, within $\bar{L}_t$, only $\frac{p_1(\vx|\cF_t)}{p_0(\vx|\cF_t)}$ remains inside the expectation. Since
\be
\Exp_0\left[\frac{p_1(\vx|\cF_t)}{p_0(\vx|\cF_t)}J(\hat{\vx},\vx) |\cF_{t}\right] = \Exp_1\left[J(\hat{\vx},\vx) |\cF_{t}\right], \nn
\ee 
we rewrite \eqref{eq:est_sep} as
\be
\hat{\x}_T^0  = \arg\min_{\hat{\vx}} \sum_{t=0}^{\infty} \big\{ b_{00} \Exp_0\left[J(\hat{\vx},\vx) |\cF_{t}\right] + b_{10}L_t \Exp_1\left[J(\hat{\vx},\vx) |\cF_{t}\right] \big\} \ind{T=t}. \nn
\ee

Define a new probability distribution 
$$
\bar{p}_t^0(\vx|\cF_t) \triangleq \frac{b_{00}p_0(\vx|\cF_t) + b_{10}L_t p_1(\vx|\cF_t)}{b_{00}+b_{10}L_t}.
$$
We are, in fact, searching for an estimator that minimizes $\bExp_0\left[J(\hat{\vx}_{T}^0,\vx) |\cF_{t}\right]$ under $\bar{p}_t^0(\vx|\cF_t)$, i.e.,
\be
\label{eq:est0}
    \hat{\x}_T^0 = \arg\min_{\hat{\vx}} \bExp_0\left[J(\hat{\vx},\vx) |\cF_T \right].
\ee
Defining the probability distribution 
$$
\bar{p}_t^1(\vx|\cF_t) \triangleq \frac{b_{01}p_0(\vx|\cF_t) + b_{11}L_t p_1(\vx|\cF_t)}{b_{01}+b_{11}L_t}
$$ 
and the expectation $\bExp_1$ for it, we can similarly show that
\be
\label{eq:est1}
    \hat{\x}_T^1 = \arg\min_{\hat{\vx}} \bExp_1\left[J(\hat{\vx},\vx) |\cF_T \right],
\ee
which, together with \eqref{eq:est0}, gives \eqref{eq:est}. The MMSE estimator, for which $J(\hat{\vx},\vx)=\|\hat{\vx}-\vx\|^2$, is given by the conditional mean $\bExp_i[\vx|\cF_T]$, hence the result in \eqref{eq:est_mmse}, concluding the proof.
\end{IEEEproof}

We see that the MMSE estimator in \eqref{eq:est_mmse} is the weighted average of the MMSE estimators under $\Hyp_0$ and $\Hyp_1$. Note that typically the likelihood ratio $L_T$ is smaller than $1$ under $\Hyp_0$ and larger than $1$ under $\Hyp_1$, that is, $\hat{\x}_T^i$ is close to $\Exp_i[\vx|\cF_T]$.

With the optimum estimators given in \eqref{eq:est} the cost function in \eqref{eq:cost} becomes
\begin{multline}
\label{eq:cost2}
	\sC(T,d_{T}) = a_0 \Pro_0(d_{T}=1|\cH_{T}) + a_1 \Pro_1(d_{T}=0|\cH_{T}) \\
+ b_{00} \Exp_0\bigg[ \underbrace{\Exp_0\left[J(\hat{\x}_{T}^0,\vx) |\cF_T\right]}_{\Delta_T^{00}} \ind{d_{T}=0}|\cH_{T}\bigg] + b_{01} \Exp_0\bigg[ \underbrace{\Exp_0\left[J(\hat{\x}_{T}^1,\vx) |\cF_T\right]}_{\Delta_T^{01}} \ind{d_{T}=1}|\cH_{T}\bigg] \\
+ b_{10} \Exp_1\bigg[ \underbrace{\Exp_1\left[J(\hat{\x}_{T}^0,\vx) |\cF_T\right]}_{\Delta_T^{10}} \ind{d_{T}=0}|\cH_{T}\bigg] + b_{11} \Exp_1\bigg[ \underbrace{\Exp_1\left[J(\hat{\x}_{T}^1,\vx) |\cF_T\right]}_{\Delta_T^{11}} \ind{d_{T}=1} |\cH_{T}\bigg],
\end{multline}
where $\Delta_T^{ij}$ is the posterior expected estimation cost when $\Hyp_j$ is decided under $\Hyp_i$.

Specifically, for the MMSE estimator
\begin{align}
\label{eq:post_cost}
\Delta_T^{ij} &= \Exp_i\left[\|\vx-\hat{\x}_T^j\|^2 |\cF_T\right] = \Exp_i\left[\|\vx-\Exp_i[\vx|\cF_T]+\Exp_i[\vx|\cF_T]-\hat{\x}_T^j\|^2 |\cF_T\right]\nn\\
&= \Exp_i\left[\|\vx-\Exp_i[\vx|\cF_T]\|^2 |\cF_T\right] + \Exp_i\left[\|\Exp_i[\vx|\cF_T]-\hat{\x}_T^j\|^2 |\cF_T\right] - 2 \Exp_i\left[ (\vx-\Exp_i[\vx|\cF_T])'(\Exp_i[\vx|\cF_T]-\hat{\x}_T^j) |\cF_T \right] \nn\\
&= \Exp_i\left[\|\vx-\Exp_i[\vx|\cF_T]\|^2 |\cF_T\right] + \|\Exp_i[\vx|\cF_T]-\hat{\x}_T^j\|^2, \\ \label{eq:post_cost1}
&= \Tr\left( \Cov_i[\vx|\cF_T] \right) + \delta_T^{ij} \| \Exp_0[\vx|\cF_T]-\Exp_1[\vx|\cF_T] \|^2,
\end{align}
where $\Tr(\cdot)$ is the trace of a matrix, 
\be
\label{eq:del_weight}
\delta_T^{0j} = \left(\frac{b_{1j}L_T}{b_{0j}+b_{1j}L_T}\right)^2
~~~\text{and}~~ \delta_T^{1j} = \left(\frac{b_{0j}}{b_{0j}+b_{1j}L_T}\right)^2. 
\ee
We used the fact that $\Exp_i[\vx|\cF_T]$ and $\hat{\x}_T^j$ are $\cF_T$-measurable, i.e., deterministic given $\cF_T$, to write \eqref{eq:post_cost}, and the MMSE estimator in \eqref{eq:est_mmse} to write \eqref{eq:post_cost1}. According to \eqref{eq:post_cost1}, $\Delta_T^{ij}$ is the MMSE under $\Hyp_i$ plus the distance between our estimator $\hat{\x}_T^j$ and the optimum estimator under $\Hyp_i$. The latter is the penalty we pay for not knowing the true hypothesis.

Note that for $b_{00}=b_{10}$ and $b_{01}=b_{11}$ (e.g., the case $b_{ij}=b~\forall i,j$, where we do not differentiate between estimation errors), the optimum estimators $\hat{\x}_T^0$ and $\hat{\x}_T^1$ in \eqref{eq:est} are both given by 
\be
\hat{\x}_T = \arg\min_{\hat{\vx}} \bExp[J(\hat{\vx},\vx)|\cF_T], \nn
\ee 
where $\bExp$ is the expectation under the distribution 
$$
\bar{p}_t(\vx|\cF_t)=\frac{p_0(\vx|\cF_t)+L_t p_1(\vx|\cF_t)}{1+L_t}.
$$ 
In particular, the MMSE estimators in \eqref{eq:est_mmse} become
\be
\hat{\x}_T=\frac{\Exp_0[\vx|\cF_T] + L_T \Exp_1[\vx|\cF_T]}{1+L_T}, \nn
\ee
regardless of the detection decision, and in \eqref{eq:post_cost1}
$$
\delta_T^{00}=\delta_T^{01}=\left(\frac{L_T}{1+L_T}\right)^2 ~~~\text{and}~~ 
\delta_T^{10}=\delta_T^{11}=\left(\frac{1}{1+L_T}\right)^2.
$$ 
\\

\subsubsection*{\underline{Optimum Detector}}

We now search for the optimum decision function $\sd_T$ that minimizes \eqref{eq:cost2} for any stopping time $T$. 

\begin{lem}
\label{lem:det}
The optimum detector $\sd_T$ for the problem in \eqref{eq:pro} is given by
\be
\label{eq:dec_opt}
    \sd_T = \left\{ \ba{ll} 1 &\text{if}~~ L_T\left(a_1 + b_{10}\Delta_T^{10} - b_{11}\Delta_T^{11}\right) \geq a_0 + b_{01}\Delta_T^{01} - b_{00}\Delta_T^{00} \\ 0 &\text{otherwise} \ea \right.,
\ee
where $L_T=\frac{p_1(\{\vy_t\}_{t=1}^T|\cH_T)}{p_0(\{\vy_t\}_{t=1}^T|\cH_T)}$ is the likelihood ratio, and $\Delta_T^{ij}=\Exp_i\left[J(\hat{\x}_{T}^j,\vx) |\cF_T\right]$ is the posterior expected estimation cost.
\end{lem}

\begin{IEEEproof}
The expectation in the definition of $\Delta_T^{ij}$ is with respect to $\vx$ only, i.e., $\Delta_T^{ij}$ is a function of the observations $\{(\vy_1,\mH_1),\ldots,(\vy_T,\mH_T)\}$ only. Similarly, the decision function $d_T$ is a function of $\{(\vy_1,\mH_1),\ldots,(\vy_T,\mH_T)\}$ only. Since, in \eqref{eq:cost2}, the probabilities in the detection cost, and the expectations in the estimation cost are conditional on $\{\mH_1,\ldots,\mH_T\}$, they are with respect to $\{\vy_1,\ldots,\vy_T\}$ only. 

Hence, in \eqref{eq:cost2}, using the likelihood ratio $L_T$ we can change the probability measure as 
\begin{align}
\Pro_1(d_T=0|\cH_T) &=  \Pro_0(L_T d_T=0|\cH_T), \nn\\
\text{and}~~ \Exp_1[\Delta_T^{1j}\ind{d_T=j}|\cH_T] &= \Exp_0[L_T \Delta_T^{1j}\ind{d_T=j}|\cH_T],~~j=0,1, \nn
\end{align} 
and combine all the terms under $\Exp_0$, i.e.,
\begin{multline}
	\sd_T = \arg \min_{d_T} \Exp_0\left[ a_0\ind{d_T=1} + a_1 L_T\ind{d_T=0} + b_{00}\Delta_T^{00} \ind{d_T=0} + b_{01}\Delta_T^{01} \ind{d_T=1} \right. \\ \left. + b_{10}L_T\Delta_T^{10} \ind{d_T=0} + b_{11}L_T \Delta_T^{11} \ind{d_T=1} | \cH_T \right], \nn
\end{multline}
where we used $\Pro(\cdot)=\Exp[\ind{\cdot}]$. Since $\ind{d_T=0}=1-\ind{d_T=1}$,
\begin{multline}
\label{eq:dec2}
	\sd_T = \arg \min_{d_T} \Exp_0\left[ \left\{ a_0 + b_{01}\Delta_T^{01} - b_{00}\Delta_T^{00} - \left(a_1 + b_{10}\Delta_T^{10} - b_{11}\Delta_T^{11}\right) L_T \right\}\ind{d_T=1} | \cH_T \right] \\ + a_1 + b_{00}\Exp_0[\Delta_T^{00}|\cH_T] + b_{10}\Exp_1[\Delta_T^{10}|\cH_T].
\end{multline}
Note that 
$$
a_1 + b_{00}\Exp_0[\Delta_T^{00}|\cH_T] + b_{10}\Exp_1[\Delta_T^{10}|\cH_T]
$$ 
does not depend on $d_T$, and the term inside the first expectation is minimized by \eqref{eq:dec_opt}. More specifically, the indicator function $\ind{d_T=1}$ is the minimizer when it only passes the negative values of the term inside the curly braces in \eqref{eq:dec2}.
\end{IEEEproof}

The optimum decision function $\sd_t$ is coupled with the estimators $\hat{\x}_t^0$, $\hat{\x}_t^1$ through the posterior estimation costs $\{\Delta_t^{ij}\}$ due to our joint formulation [cf. \eqref{eq:cost}]. Specifically, while making a decision, it takes into account, in a very intuitive way, all possible estimation costs that may result from the true hypothesis and its decision. For example, under $\Hyp_1$ small $\Delta_T^{11}$, which is the estimation cost for deciding on $\Hyp_1$, facilitates satisfying the inequality in \eqref{eq:dec_opt}, and thus favors $\sd_T=1$. Similarly, small $\Delta_T^{00}$ favors $\sd_T=0$. On the other hand, the reverse is true for $\Delta_T^{10}$ and $\Delta_T^{01}$, which correspond to the wrong decision cases. That is, large $\Delta_T^{ij}$, the cost for deciding $\Hyp_j$ under $\Hyp_i,~i\not=j$, favors $\sd_T=i$.
In the detection-only problem with $b_{ij}=0, \forall i,j$, the coupling disappears, and $\sd_T$ boils down to the well-known likelihood ratio test.\\

\subsubsection*{\underline{Complete Solution}}

We can now identify the optimum stopping time $\T$, and as a result the complete solution $(\T,\sd_\T,\hat{\x}_\T^0,\hat{\x}_\T^1)$ to the optimization problem in \eqref{eq:pro}.
\begin{thm}
\label{thm:general}
  The optimum sequential joint detector and estimator $(\T,\sd_\T,\hat{\x}_\T^0,\hat{\x}_\T^1)$ that solves the problem in \eqref{eq:pro} is given by
  \begin{align}
  \label{eq:opt_stop}
    \T &= \min\{ t\in\bN:~ \cC_t \leq \alpha \} \\
    \label{eq:opt_dec}
    \sd_\T &= \left\{ \ba{ll} 1 &\text{if}~~ L_\T\left(a_1 + b_{10}\Delta_\T^{10} - b_{11}\Delta_\T^{11}\right) \geq a_0 + b_{01}\Delta_\T^{01} - b_{00}\Delta_\T^{00} \\ 0 &\text{otherwise.} \ea \right. \\
    \label{eq:opt_est}
    \hat{\x}_\T^i &= \arg\min_{\hat{\vx}} \bExp_i\left[J(\hat{\vx},\vx) |\cF_\T\right],~i=0,1, \\
    \Big(\text{e.g.,} ~\hat{\x}_\T^i &= \frac{b_{0i}\Exp_0[\vx|\cF_\T] + b_{1i}L_\T \Exp_1[\vx|\cF_\T]}{b_{0i}+b_{1i}L_\T} ~~~\text{for}~~ J(\hat{\vx},\vx)=\|\hat{\vx}-\vx\|^2 \Big) \label{eq:opt_est_mse}
  \end{align}
  where
  \be
  \label{eq:opt_cost}
    \cC_t \triangleq \Exp_0\left[\left\{ a_0 + b_{01}\Delta_t^{01} - b_{00}\Delta_t^{00} - \left(a_1 + b_{10}\Delta_t^{10} - b_{11}\Delta_t^{11}\right) L_t \right\}^- + a_1+ b_{00}\Delta_t^{00} + b_{10}L_t\Delta_t^{10} |\cH_t \right]
  \ee
  is the optimal cost at time $t$, and $A^-=\min(A,0)$. The probability distribution $\bar{p}_t^i$ for the expectation $\bExp_i$, and the likelihood ratio $L_t$ are given in \eqref{eq:pro_mix} and \eqref{eq:like}, respectively. For the posterior estimation cost $\Delta_t^{ij}$ see \eqref{eq:cost2}--\eqref{eq:del_weight}.
\end{thm}
\begin{IEEEproof}
  In Lemma \ref{lem:est}, we showed that $\hat{\x}_T^0$ and $\hat{\x}_T^1$ minimize the cost function in \eqref{eq:cost} for any stopping time $T$ and decision function $d_T$, i.e.,
  $
    \sC(T,d_T,\hat{\x}_T^0,\hat{\x}_T^1) \leq \sC(T,d_{T},\hat{\vx}_{T}^0,\hat{\vx}_{T}^1).
  $
  Later in Lemma \ref{lem:det}, we showed that $\sC(T,\sd_T,\hat{\x}_T^0,\hat{\x}_T^1) \leq \sC(T,d_T,\hat{\x}_T^0,\hat{\x}_T^1)$. Hence, from \eqref{eq:pro}, the optimum stopping time is the first time $\cC_t \triangleq \sC(t,\sd_t,\hat{\x}_t^0,\hat{\x}_t^1)$ achieves the target accuracy level $\alpha$, as shown in \eqref{eq:opt_stop}. Since $\ind{\sd_t=1}$ filters out the positive values of
$$
a_0 + b_{01}\Delta_t^{01} - b_{00}\Delta_t^{00} - \left(a_1 + b_{10}\Delta_t^{10} - b_{11}\Delta_t^{11}\right) L_t,
$$ 
from \eqref{eq:dec2}, we write the optimal cost $\cC_t$ as in \eqref{eq:opt_cost}.
\end{IEEEproof}

According to Theorem \ref{thm:general}, the optimum scheme, at each time $t$, computes $\cC_t$, given by \eqref{eq:opt_cost}, and then compares it to $\alpha$. When $\cC_t \leq \alpha$, it stops and makes a decision using \eqref{eq:opt_dec}. Finally, it estimates $\vx$ via $\hat{\x}_T^i$, given by \eqref{eq:opt_est}, if $\Hyp_i$ is decided. 

\begin{algorithm}[h!]\small
\caption{\small The procedure for the optimum joint detection \& estimation}
\label{alg:general}
\baselineskip=0.5cm
\begin{algorithmic}[1]
\STATE Initialization: $t \gets 0$, $\cC \gets \infty$
\WHILE {$\cC > \alpha$}
    \STATE $t \gets t+1$
    \STATE $L = \frac{p_1(\{\vy_s\}_{s=1}^t|\cH_t)}{p_0(\{\vy_s\}_{s=1}^t|\cH_t)}$
    \STATE $\ve_i = \Exp_i[\vx|\cF_t],~i=0,1$
    \STATE $\text{MMSE}_i = \Tr(\Cov_i[\vx|\cF_t]),~i=0,1$
    \STATE $\Delta^{0j} = \text{MMSE}_0 + \left(\frac{b_{1j}L}{b_{0j}+b_{1j}L}\right)^2 \|\ve_0-\ve_1\|^2,~ j=0,1$
    \STATE $\Delta^{1j} = \text{MMSE}_1 + \left(\frac{b_{0j}}{b_{0j}+b_{1j}L}\right)^2 \|\ve_0-\ve_1\|^2,~ j=0,1$
    \STATE Cost: $\cC$ as in \eqref{eq:opt_cost}
\ENDWHILE
\STATE Stop: $\T = t$
\IF {$L\left(a_1 + b_{10}\Delta^{10} - b_{11}\Delta^{11}\right) \geq a_0 + b_{01}\Delta^{01} - b_{00}\Delta^{00}$}
    \STATE Decide: $\sd=1$
    \STATE Estimate: $\hat{\x} = \frac{b_{01}\ve_0 + b_{11}L\ve_1}{b_{01}+b_{11}L}$
\ELSE
    \STATE Decide: $\sd=0$
    \STATE Estimate: $\hat{\x} = \frac{b_{00}\ve_0 + b_{10}L\ve_1}{b_{00}+b_{10}L}$
\ENDIF
\end{algorithmic}
\end{algorithm}

Considering the MSE as the estimation cost function a pseudo-code for this scheme is given in Algorithm \ref{alg:general}.
Since the results in Theorem \ref{thm:general} are universal in the sense that they hold for all probability distributions and system models, in Algorithm \ref{alg:general} we provide a general procedure that requires computation of some statistics (cf. lines 4,5,6,9). In specific cases, such statistics may be easily computed. However, in many cases they cannot be written in closed forms, hence intense online computations may be required to estimate them. %The fact that such computation is performed online (i.e., as new observations arrive at each time $t$) is the bottleneck of the generic algorithm given in Algorithm \ref{alg:general}.

\vspace{2mm}
{\it Remarks:} 
\begin{enumerate}
\item In the sequential detection problem, where only the binary hypothesis testing in \eqref{eq:hypo} is of interest, the classical approach of the well-known sequential probability ratio test (SPRT) \cite{Wald47} fails to provide a feasible optimum solution due to the second observed sequence $\{\mH_t\}$. More specifically, observing the pair $\{(\vy_t,\mH_t)\}$ we end
up with a two-dimensional optimal stopping problem which is impossible to solve analytically since the thresholds for the running likelihood ratio will depend on the sequence $\{\mH_t\}$. On the other hand, for $b_{ij}=0,~i,j=0,1$, i.e., in the pure detection problem, the decision function in Theorem \ref{thm:general} boils down to the well-known likelihood ratio test (LRT). Hence, for this challenging sequential detection problem, following an alternative approach we provide an optimum sequential detector, composed of LRT and the optimum stopping time given by \eqref{eq:opt_stop} with the optimal cost 
$$
\cC_t = a_1 + \Exp_0\left[\left( a_0 - a_1 L_t \right)^- |\cH_t \right].
$$
Unlike SPRT, the above sequential detector follows a two-step procedure: it first determines the stopping time using a single threshold, and then decides using another threshold. Whereas, in SPRT, two thresholds are used in a single-step procedure to both stop and decide. 

\vspace{2mm}
\item The optimum scheme given by Theorem \ref{thm:general} is considerably more general and different than the one presented in \cite{Yilmaz_jde}. Firstly, the estimator here is the optimum estimator under a weighted average of the probability distributions under $\Hyp_0$ and $\Hyp_1$ since there are unknown parameter vectors under both hypotheses. The weights for the estimator [see \eqref{eq:pro_mix}] depend on the likelihood ratio $L_t$, hence the detector. That is, the optimum estimator for the general problem introduced in \eqref{eq:sig_nonlin}--\eqref{eq:pro} is coupled with the optimum detector. Whereas, no such coupling exists for the estimator in \cite{Yilmaz_jde}, which is the optimum estimator under $\Hyp_1$ as the unknown parameter appears only under $\Hyp_1$ ($x=0$ under $\Hyp_0$). Secondly, the optimum detector in \eqref{eq:opt_dec} is coupled with the estimator through the posterior estimation cost $\Delta_\T^{ij}$ under the four combinations of the true and selected hypotheses. On the other hand, the optimum detector in \cite{Yilmaz_jde} uses the estimator itself, which is a special case of the detector in \eqref{eq:opt_dec}. Specifically, with $b_{01}=b_{00}=0$ and $b_{10}=b_{11}$, the optimum estimator is given by $\hat{\mathsf{x}}_\T=\Exp_1[x|\cF_\T]$ when $\Hyp_1$ is decided ($x=0$ under $\Hyp_0$, hence $\hat{\mathsf{x}}_\T=0$ when $\Hyp_0$ is decided), and accordingly $\Delta_\T^{11}=\Var_1[x|\cF_\T]$, $\Delta_\T^{10}=\Var_1[x|\cF_\T]+\hat{\mathsf{x}}_\T^2$. Substituting these terms in \eqref{eq:opt_dec} we obtain the detector in \cite[Lemma 2]{Yilmaz_jde}.
Moreover, the scheme presented in Theorem \ref{thm:general} is optimum for a general non-linear model with arbitrary cost function $J(\hat{\vx},\vx)$, noise distribution, and number of parameters; and it covers the optimum scheme in \cite{Yilmaz_jde} as a special case. In \cite{Yilmaz_jde}, a monotonicity feature that facilitates the computation of the optimum stopping time is shown after a quite technical proof. Although such a monotonicity feature cannot be shown here due to the generic model we use, the optimum stopping time is still found through numerical procedures. 
\end{enumerate}

\subsection{Separated Detection and Estimation Costs}
\label{sec:sep}

In the combined cost function, given by \eqref{eq:cost}, if we penalize the wrong decisions only with the detection costs, i.e., $b_{01}=b_{10}=0$, we get the following simplified alternative cost function
\begin{multline}
\label{eq:cost_simp}
	\sC(T,d_{T},\hat{\vx}_{T}^0,\hat{\vx}_{T}^1) = a_0 \Pro_0(d_{T}=1|\cH_{T}) + a_1 \Pro_1(d_{T}=0|\cH_{T}) \\
+ b_{00} \Exp_0\left[J(\hat{\vx}_{T}^0,\vx) \ind{d_{T}=0}|\cH_{T}\right] + b_{11} \Exp_1\left[J(\hat{\vx}_{T}^1,\vx) \ind{d_{T}=1} |\cH_{T}\right].
\end{multline}
In this alternative form, detection and estimation costs are used to penalize separate cases. Specifically, under $\Hyp_i$, the wrong decision case is penalized with the constant detection cost $a_i$, and the correct decision case is penalized with the estimation cost $\Exp_i[J(\hat{\vx}_{T}^i,\vx)|\cH_t]$. Since $a_i$ is the only cost to penalize the wrong decision case, it is typically assigned a larger number here than in \eqref{eq:cost}.

The optimum scheme is obtained by substituting $b_{01}=b_{10}=0$ in Theorem \ref{thm:general}.

\begin{cor}
\label{cor:sep}
Considering the combined cost function with separated detection and estimation costs, given by \eqref{eq:cost_simp},
the optimum sequential joint detector and estimator $(\T,\sd_\T,\hat{\x}_\T^0,\hat{\x}_\T^1)$ for the problem in \eqref{eq:pro} is given by
  \begin{align}
  \label{eq:stop_simp}
    \T &= \min\{ t\in\bN:~ \cC_t \leq \alpha \} \\
    \label{eq:dec_simp}
    \sd_\T &= \left\{ \ba{ll} 1 &\text{if}~~ L_\T\left(a_1 - b_{11}\Delta_\T^{11}\right) \geq a_0 - b_{00}\Delta_\T^{00} \\ 0 &\text{otherwise.} \ea \right. \\
    \label{eq:est_simp}
    \hat{\x}_\T^i &= \arg\min_{\hat{\vx}} \Exp_i\left[J(\hat{\vx},\vx) |\cF_\T\right],~i=0,1, \\
%\label{eq:est_simp_eg}
	\Big(\text{e.g.,}~\hat{\x}_\T^i &= \Exp_i[\vx|\cF_\T] ~~~\text{for}~~ J(\hat{\vx},\vx)=\|\hat{\vx}-\vx\|^2 \Big), \nn
  \end{align}
  where
  \be
\label{eq:sep_cost}
  \cC_t = \Exp_0\left[\left\{ a_0 - b_{00}\Delta_t^{00} - \left(a_1 - b_{11}\Delta_t^{11}\right) L_t \right\}^- + a_1+ b_{00}\Delta_t^{00} |\cH_t \right],
\ee
  is the optimal cost at time $t$.
\end{cor}

The optimum stopping time, given in \eqref{eq:stop_simp}, has the same structure as in Theorem \ref{thm:general}, with a simplified optimal cost, given in \eqref{eq:sep_cost}. 

Since here we are not interested in minimizing the estimation costs in case of wrong decisions, when we decide $\Hyp_i$, we use the optimum estimator under $\Hyp_i$ [cf. \eqref{eq:est_simp}]. Recall that in Theorem \ref{thm:general}, the optimum estimator is a mixture of the optimum estimators under both hypotheses. Consequently, the posterior expected estimation cost in the correct decision case achieves the minimum, i.e.,
\be
\label{eq:post_cost_simp}
    \Delta_\T^{ii}=\min_{\hat{\vx}} \Exp_i[J(\hat{\vx},\vx)|\cF_\T]. \nn
\ee
For the MSE criterion, with $J(\hat{\x},\vx)=\|\hat{\x}-\vx\|^2$,
\be
\label{eq:post_cost_simp_mse}
    \Delta_\T^{ii}=\Tr(\Cov_i[\vx|\cF_\T])=\text{MMSE}_{\T,i}. \nn
\ee
On the other hand, in the wrong decision case, which is not of interest here, the posterior estimation cost $\Delta_\T^{ij},~i\not=j,$ is higher than that in Theorem \ref{thm:general}. 

The optimum detector in \eqref{eq:dec_simp} is biased towards the hypothesis with better estimation performance. For instance, when the minimum posterior estimation cost (e.g., MMSE) under $\Hyp_1$ is smaller than that under $\Hyp_0$ (i.e., $\Delta_\T^{11}<\Delta_\T^{00}$), it is easier to satisfy the inequality 
$$
L_\T\left(a_1 - b_{11}\Delta_\T^{11}\right) \geq a_0 - b_{00}\Delta_\T^{00},
$$
and thus to decide in favor of $\Hyp_1$. Conversely, $\Hyp_0$ is favored when $\Delta_\T^{00}<\Delta_\T^{11}$. Considering the MSE estimation cost we can call it {\em ML \& MMSE detector} since it uses the maximum likelihood (ML) criterion, as in the likelihood ratio test, together with the MMSE criterion.

\section{Linear Quadratic Gaussian (LQG) Model}
\label{sec:lin}

In this section, we consider the commonly used linear quadratic Gaussian (LQG) model, as a special case. In particular, we have the quadratic (i.e., MSE) estimation cost 
\be
\label{eq:MSE_cost}
J(\hat{\vx},\vx)=\|\hat{\vx}-\vx\|^2,
\ee
and the linear system model
\be
\label{eq:linear}
    \vy_t = \mH_t \vx + \vw_t,
\ee
where $\mH_t\in\bR^{M \times N}$, $\vw_t$ is the white Gaussian noise with covariance $\sigma^2\mI$, and $\vx$ is Gaussian under both hypotheses, i.e.,
\begin{align}
\label{eq:hypo_lqg}
\begin{split}
	\Hyp_0&:  \vx\sim \cN(\vmu_0,\mSi_0), \\
	\Hyp_1&:  \vx\sim \cN(\vmu_1,\mSi_1).
\end{split}
\end{align}

We next derive the closed-form expressions for the sufficient statistics for the optimum scheme presented in Theorem \ref{thm:general}. Using \eqref{eq:cond_mean_lqg}--\eqref{eq:Uv_def}, the optimum stopping time, detector, and estimator can be computed as in \eqref{eq:opt_stop}, \eqref{eq:opt_dec}, and \eqref{eq:opt_est_mse}, respectively.

\begin{pro}
\label{pro:lqg}
Considering the LQG model in \eqref{eq:MSE_cost}--\eqref{eq:hypo_lqg}, the sufficient statistics for the optimum sequential joint detector and estimator, presented in Theorem \ref{thm:general}, namely the conditional mean $\Exp_i[\vx|\cF_\T]$, the posterior estimation cost $\Delta_\T^{ij} = \Exp_i\left[\|\vx-\hat{\x}_\T^j\|^2 |\cF_\T\right]$ for deciding $\Hyp_j$ under $\Hyp_i$, and the likelihood ratio $L_\T=\frac{p_1(\{\vy_t\}_{t=1}^\T|\cH_\T)}{p_0(\{\vy_t\}_{t=1}^\T|\cH_\T)}$ are written as
  \begin{align}
  \label{eq:cond_mean_lqg}
    \Exp_i[\vx|\cF_\T] &= \left(\frac{\mU_\T}{\sigma^2}+\mSi_i^{-1}\right)^{-1} \left(\frac{\vv_\T}{\sigma^2}+\mSi_i^{-1}\vmu_i\right), \\
    \label{eq:Del_lqg}
    \Delta_\T^{ij} &= \Tr\left( \left(\frac{\mU_\T}{\sigma^2}+\mSi_i^{-1}\right)^{-1} \right) + \delta_\T^{ij} \| \Exp_0[\vx|\cF_\T]-\Exp_1[\vx|\cF_\T] \|^2,     
  \end{align}
  \begin{multline}
\label{eq:like_lqg}
    L_\T = \sqrt{ \frac{|\mSi_0|\left|\frac{\mU_\T}{\sigma^2}+\mSi_0^{-1}\right|} {|\mSi_1|\left|\frac{\mU_\T}{\sigma^2}+\mSi_1^{-1}\right|}}
    ~\exp\bigg[ \frac{1}{2} \bigg( \left\|\frac{\vv_\T}{\sigma^2}+\mSi_1^{-1}\vmu_1\right\|^2_{\left(\frac{\mU_\T}{\sigma^2}+\mSi_1^{-1}\right)^{-1}} - \left\|\frac{\vv_\T}{\sigma^2}+\mSi_0^{-1}\vmu_0\right\|^2_{\left(\frac{\mU_\T}{\sigma^2}+\mSi_0^{-1}\right)^{-1}}\\
    + \|\vmu_0\|^2_{\mSi_0^{-1}} - \|\vmu_1\|^2_{\mSi_1^{-1}} \bigg)  \bigg],
\end{multline}
where $\|\vx\|^2_{\mSi}\triangleq\vx'\mSi\vx$,
\be
\label{eq:Uv_def}
\mU_\T \triangleq \sum_{t=1}^\T \mH_t'\mH_t, ~~~~~ \vv_\T \triangleq \sum_{t=1}^\T \mH_t'\vy_t,
\ee
\be
\delta_\T^{0j} = \left(\frac{b_{1j}L_\T}{b_{0j}+b_{1j}L_\T}\right)^2
~~~\text{and}~~ \delta_\T^{1j} = \left(\frac{b_{0j}}{b_{0j}+b_{1j}L_\T}\right)^2. \nn
\ee
\end{pro}

\begin{IEEEproof}
We start by deriving the joint distribution density function of $\{\vy_s\}_{s=1}^t$ and $\vx$ as follows:
\begin{align}\label{eq:jointdist_xy}
&p_i(\{\vy_s\}_{s=1}^t,\vx|\cH_t)=p_i(\{\vy_s\}_{s=1}^t|\vx,\cH_t)p_i(\vx)\nonumber\\&=\frac{\exp\left(-\frac{1}{2\sigma^2}\sum_{s=1}^t \|\vy_s-\mH_s\vx\|^2\right)}{(2\pi)^{mt/2}\sigma^{mt}}~\frac{\exp\left( -\frac{1}{2}\|\vx-\vmu_i\|^2_{\mSi_i^{-1}} \right)}{(2\pi)^{n/2}|\mSi_i|^{1/2}}\nonumber\\
&=\exp\left(-\frac{1}{2}\left( \sum_{s=1}^t\frac{\|\vy_s\|^2}{\sigma^2} + \|\vmu_i\|^2_{\mSi_i^{-1}} - \|\frac{\vv_t}{\sigma^2}+\mSi_i^{-1}\vmu_i\|^2_{\left(\frac{\mU_t}{\sigma^2}+\mSi_i^{-1}\right)^{-1}} \right)\right)\nonumber\\& \exp\left( -\frac{1}{2} \left\| \vx - \left(\frac{\mU_t}{\sigma^2}+\mSi_i^{-1}\right)^{-1} \left(\frac{\vv_t}{\sigma^2}+\mSi_i^{-1}\vmu_i\right) \right\|^2_{\frac{\mU_t}{\sigma^2}+\mSi_i^{-1}} \right)\frac{1}{(2\pi)^{mt/2+n/2}\sigma^{mt}|\mSi_i|^{1/2}},
\end{align}
where $\mU_t = \sum_{s=1}^t \mH_s'\mH_s$ and $\vv_t = \sum_{s=1}^t \mH_s'\vy_s$.
Recalling that the Gaussian prior $p_i(\vx)$ is a conjugate prior for the Gaussian likelihood function $p_i(\{\vy_s\}_{s=1}^t|\vx,\cH_t)$, thus the posterior distribution $p_i(\vx|\{\vy_s\}_{s=1}^t, \cH_t)$ is also a Gaussian distribution. Furthermore, due to $$
p_i(\vx|\cF_t)=\frac{p_i(\{\vy_s\}_{s=1}^t,\vx|\cH_t)}{p_i(\{\vy_s\}_{s=1}^t|\cH_t)},
$$ we can read off the mean and variance of $\vx|\cF_t$ from the second exponent in \eqref{eq:jointdist_xy}, which is the only term involving $\vx$ in $p_i(\vx|\cF_t)$,  and arrive at
\be
\label{eq:gaus_dist}
     \vx|\cF_t \sim \cN\Bigg( \underbrace{\left(\frac{\mU_t}{\sigma^2}+\mSi_i^{-1}\right)^{-1} \left(\frac{\vv_t}{\sigma^2}+\mSi_i^{-1}\vmu_i\right)}_{\Exp_i[\vx|\cF_t]}, \underbrace{\left(\frac{\mU_t}{\sigma^2}+\mSi_i^{-1}\right)^{-1}}_{\Cov_i[\vx|\cF_t]} \Bigg),
\ee
which proves \eqref{eq:cond_mean_lqg}. Moreover, \eqref{eq:post_cost1} and \eqref{eq:gaus_dist} give \eqref{eq:Del_lqg}.
Finally, the likelihood function of $\{\vy_s\}_{s=1}^t$ is computed as
\begin{align}
\label{eq:gaus1}
    p_i(\{\vy_s\}_{s=1}^t|\cH_t) &= \frac{p_i(\{\vy_s\}_{s=1}^t,\vx|\cH_t)}{p_i(\vx|\{\vy_s\}_{s=1}^t)}\nonumber\\
    &= \frac{\exp\left[-\frac{1}{2}\left( \sum_{s=1}^t\frac{\|\vy_s\|^2}{\sigma^2} + \|\vmu_i\|^2_{\mSi_i^{-1}} - \|\frac{\vv_t}{\sigma^2}+\mSi_i^{-1}\vmu_i\|^2_{\left(\frac{\mU_t}{\sigma^2}+\mSi_i^{-1}\right)^{-1}} \right)\right]} {(2\pi)^{mt/2}~\sigma^{mt}~|\mSi_i|^{1/2}~\left|\frac{\mU_t}{\sigma^2}+\mSi_i^{-1}\right|^{1/2}}.
\end{align}
The likelihood ratio $L_\T$ in \eqref{eq:like_lqg} follows from \eqref{eq:gaus1}, concluding the proof.
\ignore{
We start with deriving the likelihood $p_i(\{\vy_s\}_{s=1}^t|\cH_t)$, which plays a key role in proving \eqref{eq:cond_mean_lqg}--\eqref{eq:like_lqg}. 
\begin{multline}
\label{eq:gaus1}
    p_i(\{\vy_s\}_{s=1}^t|\cH_t) = \int_{\bR^N} p_i(\{\vy_s\}_{s=1}^t,\vx|\cH_t) ~\text{d}\vx \\
    = \int_{\bR^N} \underbrace{\frac{\exp\left(-\frac{1}{2\sigma^2}\sum_{s=1}^t \|\vy_s-\mH_s\vx\|^2\right)}{(2\pi)^{mt/2}\sigma^{mt}}}_{p_i(\{\vy_s\}_{s=1}^t|\cH_t,\vx)} ~\underbrace{\frac{\exp\left( -\frac{1}{2}\|\vx-\vmu_i\|^2_{\mSi_i^{-1}} \right)}{(2\pi)^{n/2}|\mSi_i|^{1/2}}}_{\pi_i(\vx)}  ~\text{d}\vx \\
    = \frac{\exp\left[-\frac{1}{2}\left( \sum_{s=1}^t\frac{\|\vy_s\|^2}{\sigma^2} + \|\vmu_i\|^2_{\mSi_i^{-1}} - \|\frac{\vv_t}{\sigma^2}+\mSi_i^{-1}\vmu_i\|^2_{\left(\frac{\mU_t}{\sigma^2}+\mSi_i^{-1}\right)^{-1}} \right)\right]} {(2\pi)^{mt/2}~\sigma^{mt}~|\mSi_i|^{1/2}~\left|\frac{\mU_t}{\sigma^2}+\mSi_i^{-1}\right|^{1/2}}\\
    \underbrace{\bigintss_{\bR^N}
    \frac{\exp\left( -\frac{1}{2} \left\| \vx - \left(\frac{\mU_t}{\sigma^2}+\mSi_i^{-1}\right)^{-1} \left(\frac{\vv_t}{\sigma^2}+\mSi_i^{-1}\vmu_i\right) \right\|^2_{\frac{\mU_t}{\sigma^2}+\mSi_i^{-1}} \right)}
    {(2\pi)^{n/2} ~\left|\left(\frac{\mU_t}{\sigma^2}+\mSi_i^{-1}\right)^{-1}\right|^{1/2}}  ~\text{d}\vx}_{=1},
\end{multline}
where $\mU_t = \sum_{s=1}^t \mH_s'\mH_s$ and $\vv_t = \sum_{s=1}^t \mH_s'\vy_s$. The likelihood ratio $L_\T$ in \eqref{eq:like_lqg} follows from \eqref{eq:gaus1}. 
We see that the joint distribution $p_i(\{\vy_s\}_{s=1}^t,\vx|\cH_t)$ is given by \eqref{eq:gaus1} without the integral sign. Hence, the posterior distribution 
$$
p_i(\vx|\cF_t)=\frac{p_i(\{\vy_s\}_{s=1}^t,\vx|\cH_t)}{p_i(\{\vy_s\}_{s=1}^t|\cH_t)}
$$ 
equals the term inside the integral in \eqref{eq:gaus1}, i.e., 
\be
\label{eq:gaus_dist}
     \vx|\cF_t \sim \cN\Bigg( \underbrace{\left(\frac{\mU_t}{\sigma^2}+\mSi_i^{-1}\right)^{-1} \left(\frac{\vv_t}{\sigma^2}+\mSi_i^{-1}\vmu_i\right)}_{\Exp_i[\vx|\cF_t]}, \underbrace{\left(\frac{\mU_t}{\sigma^2}+\mSi_i^{-1}\right)^{-1}}_{\Cov_i[\vx|\cF_t]} \Bigg),
\ee
which proves \eqref{eq:cond_mean_lqg}. Moreover, \eqref{eq:post_cost1} and \eqref{eq:gaus_dist} give \eqref{eq:Del_lqg}, concluding the proof.}
\end{IEEEproof}

Note that the sufficient statistics in \eqref{eq:cond_mean_lqg}--\eqref{eq:like_lqg} are functions of $\mU_\T$ and $\vv_\T$ only, which are given in \eqref{eq:Uv_def}. As a result, from \eqref{eq:opt_cost}, given $\cH_t$, the expectation in the optimal cost $\cC_t$ is conditional on $\mU_t$ as $\mU_t$ is $\cH_t$-measurable, and hence the expectation is taken over $\vv_t$. That is, $\cC_t$ and the optimum stopping time $\T$, given by \eqref{eq:opt_cost} and \eqref{eq:opt_stop}, respectively, are functions of $\mU_t$ only, which is in fact the Fisher information matrix scaled by $\sigma^2$. 

Using \eqref{eq:linear} and \eqref{eq:Uv_def} we can write 
$$
\vv_t = \mU_t \vx + \sum_{s=1}^t \mH_s' \vw_s,
$$ 
which is distributed as $\cN(\mU_t \vmu_i,\mU_t\mSi_i\mU_t+\sigma^2\mU_t)$ under $\Hyp_i$. At each time $t$, for the corresponding $\mU_t$, we can estimate the optimal cost $\cC_t$ through Monte Carlo simulations, and stop if $\cC_t\le\alpha$ according to \eqref{eq:opt_stop}. Specifically, given $\mU_t$ we generate realizations of $\vv_t$, compute the expression inside the expectation in \eqref{eq:opt_cost} using \eqref{eq:cond_mean_lqg}--\eqref{eq:like_lqg}, and average them. Alternatively, $\cC(\mU)$ can be computed in the same way through offline Monte Carlo simulations on a grid of $\mU$. Then, at each time $t$, checking the $\cC(\mU^*)$ value for the average $\mU^*$ of $2^{\frac{N^2+N}{2}}$ neighboring points to $\mU_t$ (or simply the closest grid point $\mU^*$ to $\mU_t$) we can decide to stop if $\cC(\mU^*)\le\alpha$ or to continue if $\cC(\mU^*)>\alpha$. Although $\mU_t=\sum_{s=1}^t \mH_s'\mH_s$ has $N^2$ entries, due to symmetry the grid for offline simulations is $\frac{N^2+N}{2}$-dimensional.

\subsection{Independent LQG Model}
\label{sec:ind}

Here, we further assume in \eqref{eq:linear} that the entries of $\vx$ are independent [i.e., $\mSi_0$ and $\mSi_1$ are diagonal in \eqref{eq:hypo_lqg}], and $\mH_t$ is diagonal. Note that in this case $M=N$, and the entries of $\vy_t$ are independent. This may be the case in a distributed system (e.g., wireless sensor network) in which each node (e.g., sensor) takes noisy measurements of a local parameter, and there is a global event whose occurrence changes the probability distributions of local parameters. In such a setup, nodes collaborate through a fusion center to jointly detect the global event and estimate the local parameters. To find the optimal scheme we assume that all the observations collected at nodes are available to the fusion center.

\begin{pro}
\label{pro:ind}
Considering the independent LQG model with diagonal $\mH_t$ and $\mSi_i$ in \eqref{eq:linear} and \eqref{eq:hypo_lqg}, respectively, the necessary and sufficient statistics for the optimum scheme in Theorem \ref{thm:general} are written as
  \begin{align}
  \label{eq:cond_mean_lqg_ind}
    \Exp_i[\vx|\cF_\T] &= [\bar{x}_1,\ldots,\bar{x}_N]', ~~~ \bar{x}_n = \frac{\frac{v_{\T,n}}{\sigma^2} + \frac{\mu_{i,n}}{\rho_{i,n}^2}} {\frac{u_{\T,n}}{\sigma^2}+\frac{1}{\rho_{i,n}^2}}, \\
    \label{eq:Del_lqg_ind}
    \Delta_\T^{ij} &= \sum_{n=1}^N \frac{1}{\frac{u_{\T,n}}{\sigma^2}+\frac{1}{\rho_{i,n}^2}} + \delta_\T^{ij} \| \Exp_0[\vx|\cF_\T]-\Exp_1[\vx|\cF_\T] \|^2,     \\
    \label{eq:Like_lqg_ind}
    L_\T &= \prod_{n=1}^N \frac{\rho_{0,n}}{\rho_{1,n}}~ \sqrt{\frac{\frac{u_{\T,n}}{\sigma^2}+\frac{1}{\rho_{0,n}^2}} {\frac{u_{\T,n}}{\sigma^2}+\frac{1}{\rho_{1,n}^2}}}~ \exp\left[ \frac{1}{2} \left( \frac{\left( \frac{v_{\T,n}}{\sigma^2} + \frac{\mu_{1,n}}{\rho_{1,n}^2} \right)^2} {\frac{u_{\T,n}}{\sigma^2}+\frac{1}{\rho_{1,n}^2}} - \frac{\left( \frac{v_{\T,n}}{\sigma^2} + \frac{\mu_{0,n}}{\rho_{0,n}^2} \right)^2} {\frac{u_{\T,n}}{\sigma^2}+\frac{1}{\rho_{0,n}^2}} + \frac{\mu_{0,n}^2}{\rho_{0,n}^2} - \frac{\mu_{1,n}^2}{\rho_{1,n}^2} \right) \right],
  \end{align}
where the subscript $n$ denotes the $n$-th entry of the corresponding vector, $\rho_{i,n}^2$ and $h_{t,n}$ are the $n$-th diagonal entries of $\mSi_i$ and $\mH_t$, respectively, 
$$
u_{\T,n}=\sum_{t=1}^\T h_{t,n}^2 ~~~\text{and}~~~ v_{\T,n}=\sum_{t=1}^\T h_{t,n} y_{t,n}.
$$
\end{pro}

\begin{IEEEproof}
Since $\mH_t$ is diagonal and both $\vx$ and $\vw_t$ have independent entries, the linear system model \eqref{eq:linear} can be decomposed into $N$ sub-systems,  i.e., $y_{t, n}=h_{t, n}x_n+w_n, \; n=1, 2, \ldots, N$, which are independent from each other. Then the posterior distribution is a scalar version of \eqref{eq:gaus_dist} for each local parameter $x_n$, i.e., 
\ignore{
$$
p_i(x_n|\cF_t^n)=\frac{p_i(\{y_{s,n}\}_{s=1}^t,x_n|\cH_t^n)}{p_i(\{y_{s,n}\}_{s=1}^t|\cH_t^n)}
$$} 
%of each local parameter $x_n$ is
\be
\label{eq:ind_dist}
 x_n|\{y_{s, n}\}_{s=1}^t \sim  \cN\left( \frac{\frac{v_{t,n}}{\sigma^2} + \frac{\mu_{i,n}}{\rho_{i,n}^2}} {\frac{u_{t,n}}{\sigma^2}+\frac{1}{\rho_{i,n}^2}}, \frac{1}{\frac{u_{t,n}}{\sigma^2}+\frac{1}{\rho_{i,n}^2}} \right),
\ee
proving \eqref{eq:cond_mean_lqg_ind}. Moreover,
%As in the proof of Proposition \ref{pro:lqg}, we derive the likelihood $p_i(\{\vy_s\}_{s=1}^t|\cH_t)$.
due to spatial independence, we have
$$
p_i(\{\vy_s\}_{s=1}^t|\cH_t)=\prod_{n=1}^N p_i(\{y_{s,n}\}_{s=1}^t|\cH_t^n),
$$
where $p_i(\{y_{s,n}\}_{s=1}^t|\cH_t^n)$ is given by the scalar version of \eqref{eq:gaus1}, i.e.,
\be
\label{eq:ind1}
    p_i(\{y_{s,n}\}_{s=1}^t|\cH_t^n) = \frac{\exp\left[-\frac{1}{2}\left( \sum_{s=1}^t \frac{y_{s,n}^2}{\sigma^2} + \frac{\mu_{i,n}^2}{\rho_{i,n}^2} - \frac{\left( \frac{v_{t,n}}{\sigma^2} + \frac{\mu_{i,n}}{\rho_{i,n}^2} \right)^2} {\frac{u_{t,n}}{\sigma^2}+\frac{1}{\rho_{i,n}^2}} \right)\right]} {(2\pi)^{t/2}~\sigma^t~\rho_{i,n}~ \sqrt{\frac{u_{t,n}}{\sigma^2}+\frac{1}{\rho_{i,n}^2}}}.
    \ignore{\underbrace{\bigints_{\bR}
    \frac{\exp\left[ -\frac{\left(x_n - \frac{\frac{v_{t,n}}{\sigma^2} + \frac{\mu_{i,n}}{\rho_{i,n}^2}} {\frac{u_{t,n}}{\sigma^2}+\frac{1}{\rho_{i,n}^2}}\right)^2}{\frac{2}{\frac{u_{t,n}}{\sigma^2}+\frac{1}{\rho_{i,n}^2}}} \right]}
    {\sqrt{\frac{2\pi}{\frac{u_{t,n}}{\sigma^2}+\frac{1}{\rho_{i,n}^2}}}} ~\text{d}x_n}_{=1}.}
\ee
The global likelihood ratio is given by the product of the local ones, i.e., $L_t = \prod_{n=1}^N L_t^n$, where, from \eqref{eq:ind1}, 
$$
L_t^n=\frac{p_1(\{y_{s,n}\}_{s=1}^t|\cH_t^n)}{p_0(\{y_{s,n}\}_{s=1}^t|\cH_t^n)}
$$ 
is written as in \eqref{eq:Like_lqg_ind}. 
\ignore{Since 
$$
p_i(\{y_{s,n}\}_{s=1}^t|\cH_t) = \int_{\bR} p_i(\{y_{s,n}\}_{s=1}^t,x_n|\cH_t) ~\text{d}x_n,
$$ 
the joint distribution $p_i(\{y_{s,n}\}_{s=1}^t,x_n|\cH_t)$ is given by \eqref{eq:ind1} without the integral sign.}  
From \eqref{eq:post_cost},
$$
\Delta_\T^{ij} = \sum_{n=1}^N \Var_i[x_n|\cF_\T^n] + \delta_\T^{ij} \| \Exp_0[\vx|\cF_\T]-\Exp_1[\vx|\cF_\T] \|^2,
$$
which, together with \eqref{eq:ind_dist}, gives \eqref{eq:Del_lqg_ind}, concludes the proof. 
\end{IEEEproof}

In this case, $\Exp_i[\vx|\cF_t]$, $\Delta_t^{ij}$, and $L_t$ are functions of $\{u_{t,n},v_{t,n}\}_{n=1}^N$ only, hence the optimal cost $\cC_t$ and the optimum stopping time $\T$, given in Theorem \ref{thm:general}, are functions of $\{u_{t,n}\}_{n=1}^N$ only. At each time $t$, given $\{u_{t,n}\}_{n=1}^N$, we can estimate $\cC_t$ through Monte Carlo simulations using 
$$
v_{t,n}\sim\cN(\mu_{i,n}u_{t,n},\rho_{i,n}^2 u_{t,n}^2+\sigma^2 u_{t,n}),
$$
and \eqref{eq:opt_cost}, \eqref{eq:cond_mean_lqg_ind}--\eqref{eq:Like_lqg_ind}; and stop when the estimated $\cC_t \le \alpha$. Alternatively, $\cC(\{u_{t,n}\})$ can be computed in the same way through offline Monte Carlo simulations on a grid of $\{u_{t,n}\}_{n=1}^N$, as discussed in the general LQG case. Note that the grid here is $N$-dimensional, which is much smaller than the $\frac{N^2+N}{2}$-dimensional grid under the general LQG model. Consequently, the alternative scheme that performs offline simulations is more viable here.

\subsection{Numerical Results}
In this subsection, we compare the proposed joint detection and estimation scheme  (SJDE) with the conventional method, which invokes the sequential detector to decide between the two hypotheses and then computes the corresponding MMSE estimate. The comparison is based on the LQG model that we have investigated in this section. In particular,  for the conventional method, the commonly adopted sequential probability ratio test (SPRT) is used, followed by an MMSE estimator. SPRT computes the log-likelihood ratio, i.e., $\log L_t$, at each sampling instant and examines whether it falls in the prescribed interval, denoted as $[-B, A]$. The stopping time and decision rule of SPRT are defined as 
\begin{align}
&\T_\text{SPRT}\triangleq \min \left\{t\in \mathbb{N}: \log L_t \in [-B, A]\right\},\\
\text{and}\quad & \sd_{\T_\text{SPRT}} = \left\{ \ba{ll} 1 &\text{if}~~ L_{\T_\text{SPRT}}\ge A, \\ 0 & \text{if}~~L_{\T_\text{SPRT}}\le -B, \ea \right.
\end{align}
where $A$ and $B$, in practice, are selected such that the target accuracy level is satisfied. In the case of LQG model, $L_t$ is given by \eqref{eq:like_lqg}. Upon the decision $\sd_{\T_\text{SPRT}}$, the corresponding MMSE estimator follows. 

\begin{figure}
\centering
\includegraphics[width=0.9\textwidth]{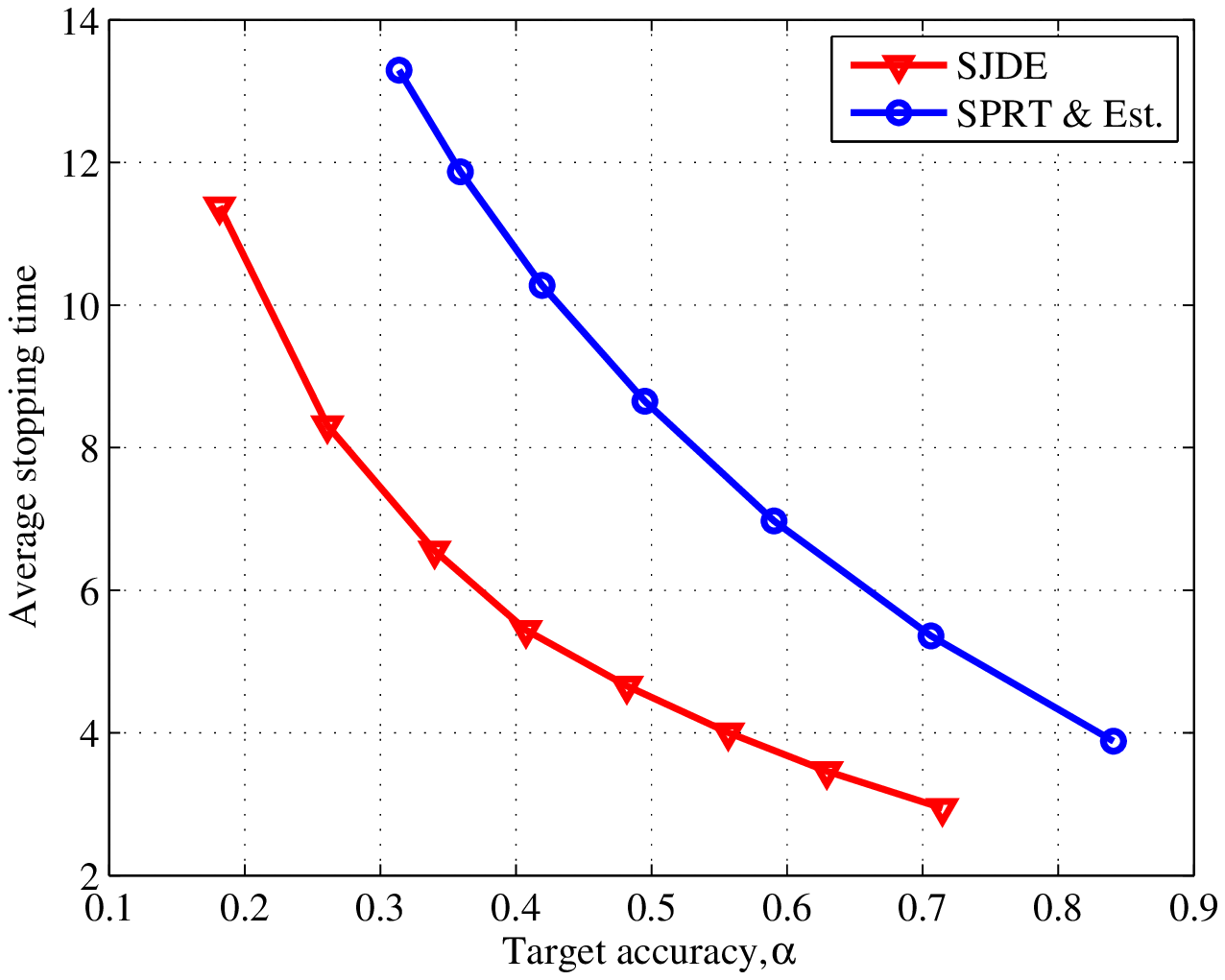}
\caption{Average stopping time vs. target accuracy level for SJDE and the combination of SPRT detector \& MMSE estimator.}\label{fig:fig3}
\end{figure}

For the numerical comparison, we consider the LQG model with ${\vx}\in \mathbb{R}^{3\times 1}$, ${\mH}_t\in \mathbb{R}^{1\times 3}$ and the following hypotheses:
\begin{align}
\begin{split}
	\Hyp_0&:  \vx\sim \cN({\bf 1},0.5{\bf I}), \\
	\Hyp_1&:  \vx\sim \cN(-{\bf 1},0.5{\bf I}),
\end{split}
\end{align}
where ${\bf 1}$ is the $3$-dimensional vector with all entries equal to $1$ and ${\bf I}$ is the identity matrix. The noise ${\vw}_t$ is white Gaussian process, ${\vw}_t\sim \mathcal{N}\left({\bf 0}, {\bf I}\right)$. $\mH_t$ is also generated as ${\mH}_t\sim \mathcal{N}\left({\bf 0}, {\bf I}\right)$ and independent over time. The parameters of the cost function are set as follows: $a_0=a_1=0.5$, $b_{00}=b_{11}=0.5$, $b_{10}=b_{01}=0$. Fig. \ref{fig:fig3} illustrates the performance of SJDE and the conventional method in terms of the average stopping time against the target accuracy, i.e, $\alpha$. Note that small $\alpha$ implies high accuracy of the detection and estimation performance, thus requiring more detection stopping time. It is seen that the SJDE (cf. red line with triangle marks) significantly outperforms the conventional combination of SPRT and MMSE (cf. blue line with circle marks). That is, SJDE exhibits a much smaller detection stopping time, while achieving the same target accuracy $\alpha$. 

%Note that, even for the composite hypothesis testing, SPRT is not optimal in terms of detection delay. 

\section{Dynamic Spectrum Access in Cognitive Radio Networks}
\label{sec:cr}

\subsection{Background}

Dynamic spectrum access is a fundamental problem in cognitive radio, in which secondary users (SUs) are allowed to utilize a wireless spectrum band (i.e., communication channel)
that is licensed to primary users (PUs) without affecting the PU quality of service (QoS) \cite{Zhao07}.
Spectrum sensing plays a key role in maximizing the SU throughput, and at the same time protecting the PU QoS. In spectrum sensing, if no PU communication is detected, then SU can opportunistically utilize the band \cite{Liang08,Chen08}. Otherwise, it has to meet some strict interference constraints. Nevertheless, it can still use the band in an underlay fashion with a transmit power that does not violate the maximum allowable interference level \cite{Kang09_1,Musavian09}.
Methods for combining the underlay and opportunistic access approaches have also been proposed, e.g., \cite{Kang09_2,Wang13,Yilmaz_dsa}. In such combined methods, the SU senses the spectrum band, as in opportunistic access, and controls its transmit power using the sensing result, which allows SU to coexist with PU, as in underlay.

The interference at the PU receiver is a result of the SU transmit power, and also the power gain of the channel between the SU transmitter and PU receiver. Hence, SU needs to estimate the channel coefficient to keep its transmit power within allowable limits. As a result, channel estimation, in addition to PU detection, is an integral part of an effective dynamic spectrum access scheme in cognitive radio. In spectrum access methods it is customary to assume perfect channel state information (CSI) at the SU, e.g., \cite{Kang09_1,Kang09_2,Musavian09}. 
It is also crucial to minimize the sensing time for maximizing the SU throughput. Specifically, decreasing the sensing period, that is used to determine the transmit power, saves time for data communication, increasing the SU throughput. Consequently, dynamic spectrum access in cognitive radio is intrinsically a sequential joint detection and estimation problem. Recently, in \cite{Yilmaz_dsa}, the joint problem of PU detection and channel estimation for SU power control has been addressed using a sequential two-step procedure. In the first step, sequential joint spectrum sensing and channel estimation is performed; and in the second stage, the SU transmit power is determined based on the results of first stage. Here, omitting the second stage, we derive the optimum scheme for the first stage in an alternative way under the general theory presented in the previous sections.

\subsection{Problem Formulation}

We consider a cognitive radio network consisting of $K$ SUs, and a pair of PUs. In PU communication, a preamble takes place before data communication for synchronization and channel estimation purposes. In particular, during the preamble both PUs transmit random pilot symbols simultaneously through full duplexing. Pilot signals are often used in channel estimation, e.g., \cite{Li00}, and also in spectrum sensing, e.g., \cite{Sahai06}.
We assume each SU observes such pilot symbols (e.g., it knows the seed of the random number generator) so that it can estimate the channels between itself and PUs. Moreover, SUs cooperate to detect the PU communication, through a fusion center (FC), which can be one of the SUs. To find the optimal scheme we assume a centralized setup where all the observations collected at SUs are available to the FC. In practice, under stringent energy and bandwidth constraints SUs can effectively report their necessary and sufficient statistics to the FC using a non-uniform sampling technique called level-triggered sampling, as proposed in \cite{Yilmaz_dsa}.

When the channel is idle (i.e., no PU communication), there is no interference constraint, and as a result SUs do not need to estimate the interference channels to determine the transmit power, which is simply the full power $P_\text{max}$. On the other hand, in the presence of PU communication, to satisfy the peak interference power constraints $I_1$ and $I_2$ of PU $1$ and PU $2$, respectively, SU $k$ should transmit with power 
$$
P_k = \min\left\{P_\text{max},\frac{I_1}{x_{1k}^2},\frac{I_2}{x_{2k}^2}\right\},
$$
where $x_{jk}$ is the channel coefficient between PU $j$ and SU $k$. Hence, firstly the presence/absence of PU communication is detected. If no PU communication is detected, then a designated SU transmits data with $P_\text{max}$. Otherwise, the channels between PUs and SUs are estimated to determine transmission powers, and then the SU with the highest transmission power starts data communication.

We can model this sequential joint detection and estimation problem using the linear model in \eqref{eq:linear}, where the vector 
$$
\vx=[x_{11},\ldots,x_{1K},x_{21},\ldots,x_{2K}]'
$$ 
holds the interference channel coefficients between PUs ($j=1,2$) and SUs ($k=1,\ldots,K$); the diagonal matrix
$$
\mH_t = \text{diag}(h_{t,1},\ldots,h_{t,1},h_{t,2},\ldots,h_{t,2}) \in \bR^{2K\times 2K}
$$
holds the PU pilot signals; and
\begin{align}
\vy_t &= [y_{t,11},\ldots,y_{t,2K}]' \nn\\
\vw_t &= [w_{t,11},\ldots,w_{t,2K}]' \nn 
\end{align} 
are the observation and Gaussian noise vectors at time $t$, respectively. Then, we have the following binary hypothesis testing problem
\begin{align}
\label{eq:hypo_cr}
\begin{split}
	\Hyp_0&:  \vx = 0, \\
	\Hyp_1&:  \vx \sim \cN(\vmu,\mSi),
\end{split}
\end{align}
where $\vmu = [\mu_{11},\ldots,\mu_{2K}]'$, $\mSi = \text{diag}(\rho_{11}^2,\ldots,\rho_{2K}^2)$ with $\mu_{jk}$ and $\rho_{jk}^2$ being the mean and variance of the channel coefficient $x_{jk}$, respectively.

Since channel estimation is meaningful only under $\Hyp_1$, we do not assign estimation cost to $\Hyp_0$, and perform estimation only when $\Hyp_1$ is decided. In other words, we use the cost function
\begin{multline}
\label{eq:cost_cr}
	\sC(T,d_{T},\hat{\vx}_{T}) = a_0 \Pro_0(d_{T}=1|\cH_{T}) + a_1 \Pro_1(d_{T}=0|\cH_{T}) \\
+ b_1 \Exp_1\left[ \|\hat{\vx}_{T}-\vx\|^2 \ind{d_{T}=1} + \|\vx\|^2 \ind{d_{T}=0} |\cH_{T}\right],
\end{multline}
which is a special case of \eqref{eq:cost}. When $\Hyp_0$ is decided, it is like we set $\hat{\vx}_T=0$. Similar to \eqref{eq:pro}, we want to solve the following problem
\be
\label{eq:pro_cr}
	\min_{T,d_{T},\hat{\vx}_{T}} T ~~ \text{s.t.} ~~ \sC(T,d_{T},\hat{\vx}_{T}) \leq \alpha,
\ee
for which the optimum solution follows from Theorem \ref{thm:general} and Proposition \ref{pro:ind}.

\subsection{Optimum Solution}

\begin{cor}
\label{cor:cr}
The optimum scheme for the sequential joint spectrum sensing and channel estimation problem in \eqref{eq:pro_cr} is given by
  \begin{align}
  \label{eq:opt_stop_cr}
    \T &= \min\{ t\in\bN:~ \cC_t \leq \alpha \} \\
    \label{eq:opt_dec_cr}
    \sd_\T &= \left\{ \ba{ll} 1 &\text{if}~~ L_\T \geq \frac{a_0}{a_1 + b_1 \|\hat{\x}_\T\|^2}  \\ 0 &\text{otherwise} \ea \right. \\
    \label{eq:opt_est_cr}
    \hat{\x}_\T &= [\bar{x}_{11},\ldots,\bar{x}_{2K}]', ~~~\text{and}~~~ \bar{x}_{jk} = \frac{\frac{v_{\T,jk}}{\sigma^2} + \frac{\mu_{jk}}{\rho_{jk}^2}} {\frac{u_{\T,j}}{\sigma^2}+\frac{1}{\rho_{jk}^2}}, 
  \end{align}
  where $u_{\T,j}=\sum_{t=1}^\T h_{t,j}^2$, $v_{\T,jk}=\sum_{t=1}^\T h_{t,j} y_{t,jk}$,
  \be
  \label{eq:opt_cost_cr}
    \cC_t \triangleq \Exp_0\left[\left\{ a_0 - \left(a_1 + b_1\|\hat{\x}_t\|^2\right) L_t \right\}^- |\cH_t \right] + b_1 \Exp_1\left[ \|\hat{\x}_t\|^2 + \sum_{j=1}^2 \sum_{k=1}^K \frac{1}{\frac{u_{t,j}}{\sigma^2}+\frac{1}{\rho_{jk}^2}} \Big|\cH_t \right] + a_1
  \ee
  is the optimal cost at time $t$; and
  \be
  \label{eq:like_cr}
  L_t = \frac{p_1(\{\vy_s\}_{s=1}^t|\cH_t)}{p_0(\{\vy_s\}_{s=1}^t|\cH_t)} = \prod_{j=1}^2 \prod_{k=1}^K \frac{\exp\left[ \frac{1}{2} \left( \frac{\left( \frac{v_{t,jk}}{\sigma^2} + \frac{\mu_{jk}}{\rho_{jk}^2} \right)^2} {\frac{u_{t,j}}{\sigma^2}+\frac{1}{\rho_{jk}^2}} - \frac{\mu_{jk}^2}{\rho_{jk}^2} \right) \right]} {\rho_{jk}~ \sqrt{\frac{u_{t,j}}{\sigma^2}+\frac{1}{\rho_{jk}^2}}}
  \ee
  is the likelihood ratio at time $t$.
\end{cor}

\begin{IEEEproof}
Substituting $b_{01}=0$ into \eqref{eq:opt_est_mse} we write the optimum estimator as
\be
\label{eq:est_cr}
    \hat{\x}_\T = \Exp_1[\vx|\cF_\T],
\ee
which is used only when $\Hyp_1$ is decided. Since the independent LQG model (i.e., diagonal $\mH_t$ and $\mSi$) is used in the problem formulation, we can borrow, from Proposition \ref{pro:ind}, the result for $\Exp_1[\vx|\cF_\T]$, given by \eqref{eq:cond_mean_lqg_ind}, to write \eqref{eq:opt_est_cr}. 

From \eqref{eq:post_cost}, we write
$$
\Delta_\T^{11}=\Tr\left( \Cov_1[\vx|\cF_T] \right) ~~~\text{and}~~~ \Delta_\T^{10}=\Tr\left( \Cov_1[\vx|\cF_T] \right) + \|\hat{\x}_\T\|^2,
$$
where we used $\hat{\x}_\T^1=\Exp_1[\vx|\cF_\T]$ and $\hat{\x}_\T^0=0$. Then, in the optimum detector expression given by \eqref{eq:opt_dec}, on the right side we only have $a_0$ since $b_{01}=b_{00}=0$; and on the left side we have $L_\T(a_1+b_1\|\hat{\x}_\T\|^2)$ since $b_{10}=b_{11}=b_1$, resulting in \eqref{eq:opt_dec_cr}.

Similarly, using $b_{01}=b_{00}=0$ and $b_{10}=b_{11}=b_1$ in \eqref{eq:opt_cost} the optimum stopping time and the optimal cost are as in \eqref{eq:opt_stop_cr} and \eqref{eq:opt_cost_cr}, respectively. For the likelihood ratio, due to independence, we have 
$$
L_t=\prod_{j=1}^2 \prod_{k=1}^K L_t^{jk} ~~~\text{where}~~~ L_t^{jk}=\frac{p_1(\{y_{s,jk}\}_{s=1}^t|\cH_t^{jk})}{p_0(\{y_{s,jk}\}_{s=1}^t|\cH_t^{jk})}
$$ 
is the local likelihood ratio for the channel between PU $j$ and SU $k$. The likelihood $p_1(\{y_{s,jk}\}_{s=1}^t|\cH_t^{jk})$ is given by \eqref{eq:ind1}; and 
$$
p_0(\{y_{s,jk}\}_{s=1}^t|\cH_t^{jk}) = \frac{\exp\left(-\frac{1}{2} \sum_{s=1}^t \frac{y_{s,jk}^2}{\sigma^2} \right)} {(2\pi)^{t/2}~\sigma^t}
$$
since the received signal under $\Hyp_0$ is white Gaussian noise. Hence, $L_t$ is written as in \eqref{eq:like_cr}.
\end{IEEEproof}

At each time $t$ the optimal cost $\cC_t$, given by \eqref{eq:opt_cost_cr}, can be estimated through
Monte Carlo simulations by generating the realizations of $v_{t,jk}$, independently for each pair $(j,k)$, according to $\cN(0,\sigma^2 u_{t,j})$ and $\cN(\mu_{jk}u_{t,j},\rho_{jk}^2 u_{t,j}^2+\sigma^2 u_{t,j})$ under $\Hyp_0$ and $\Hyp_1$, respectively. Alternatively, since $\cC_t$ is a function of $u_{t,1}$ and $u_{t,2}$ only, we can effectively estimate $\cC(u_1,u_2)$ through offline Monte Carlo simulations over the 2-dimensional grid. Note that the number of grid dimensions here is much less than $N$ and $\frac{N^2+N}{2}$ for the independent and general LQG models in Section \ref{sec:lin}, respectively. 

The optimum detector, given in \eqref{eq:opt_dec_cr}, uses the side information provided by the estimator itself. Specifically, the farther away the estimates are from zero, i.e., $\|\hat{\x}_\T\|^2 \gg 0$, the easier it is to decide for $\Hyp_1$; and the reverse is true for $\Hyp_0$. 
The optimum estimator, given by \eqref{eq:opt_est_cr}, is the MMSE estimator under $\Hyp_1$ as channel estimation is meaningful only when PU communication takes place. 

\vspace{2mm}
{\em Remark:} In \cite{Yilmaz_dsa}, following the technical proof of \cite{Yilmaz_jde} the optimum solution is presented for a similar sequential joint detection and estimation problem with complex channels. Here, under a general framework, we derive the optimum scheme following an alternative approach. Particularly, we show that, without the monotonicity property for the optimal cost, the optimum stopping time can be efficiently computed through (offline/online) Monte Carlo simulations. Furthermore, we here also show how this dynamic spectrum access method fits to the systematic theory of sequential joint detection and estimation, developed in the previous sections.

\section{State Estimation in Smart Grid with Topological Uncertainty}
\label{sec:sg}

\subsection{Background and Problem Formulation}

State estimation is a vital task in real-time monitoring of smart grid \cite{Huang12}. In the widely used linear model
\be
\label{eq:meas_sg}
	\vy_t = \mH \vx + \vw_t,
\ee
the state vector $\vx=[\theta_1,\ldots,\theta_N]'$ holds the bus voltage phase angles; the measurement matrix $\mH\in\bR^{M\times N}$ represents the network topology; $\vy_t\in\bR^M$ holds the power flow and injection measurements; and $\vw_t\in\bR^M$ is the white Gaussian measurement noise vector. We assume a pseudo-static state estimation problem, i.e., $\vx$ does not change during the estimation period. For the above linear model to be valid it is assumed that the differences between phase angles are small. Hence, we can model $\theta_n,~n=1,\ldots,N$ using a Gaussian prior with a small variance, as in \cite{Huang11,Chen13}.

The measurement matrix $\mH$ is also estimated periodically using the status data from switching devices in the power grid, and assumed to remain unchanged until the next estimation instance. However, in practice, such status data is also noisy, like the power flow measurements in \eqref{eq:meas_sg}, and thus the estimate of $\mH$ may include some error. Since the elements of $\mH$ take the values $\{-1,0,1\}$, there is a finite number of possible errors. Another source of topological uncertainty is the power outage, in which protective devices automatically isolate the faulty area from the rest of the grid. Specifically, an outage changes the grid topology, i.e., $\mH$, and also the prior on $\vx$. We model the topological uncertainty using multiple hypotheses, as in \cite{Chen13,Huang14,Zhao12,Zhao13}. In \eqref{eq:meas_sg}, under hypothesis $j$ we have
\be
    \Hyp_j:~~\mH=\mH_j,~\vx\sim\cN(\vmu_j,\mSi_j),~j=0,1,\ldots,J,
\ee
where $\Hyp_0$ corresponds to the normal-operation (i.e., no estimation error or outage) case.

Note that in this case, for large $J$, in \eqref{eq:cost} there will be a large number of cross estimation costs $b_{ji} \Exp_j[\|\hat{\vx}_T^i-\vx\|^2 \ind{d_T=i}|\cH_T],~i\not=j$ that penalize the wrong decisions under $\Hyp_j$. For simplicity, following the formulation in Section \ref{sec:sep}, we here penalize the wrong decisions only with the detection costs, i.e., $b_{ji}=0,~i\not=j$, and $b_{jj}=b_j>0$. Hence, generalizing the cost function in \eqref{eq:cost_simp} to the multi-hypothesis case, we use the following cost function
\be
\label{eq:cost_sg}
    \sC(T,d_T,\{\hat{\vx}_T^j\}) = \sum_{j=0}^J \left\{ a_j \Pro_j(d_T\not=j) + b_j\Exp_j\left[\|\hat{\vx}_T^j-\vx\|^2 \ind{d_T=j}\right] \right\}.
\ee
Here we do not need the conditioning on $\cH_t$ as the measurement matrices $\{\mH_j\}$ are deterministic and known. As a result the optimum stopping time $\T$ is deterministic and can be computed offline. %Therefore, we need to find only the optimum detector and optimum estimators.
We seek the solution to the following optimization problem,
\be
\label{eq:pro_sg}
	\min_{T,d_T,\{\hat{\vx}_T^j\}} T ~~ \text{s.t.} ~~ \sC(T,d_T,\{\hat{\vx}_T^j\}) \leq \alpha.
\ee

\subsection{Optimum Solution}

We next present the solution to \eqref{eq:pro_sg}, which includes testing of multiple hypotheses.

\begin{pro}
\label{pro:sg}
The optimum scheme for the sequential joint detection and estimation problem in \eqref{eq:pro_sg} is given by
  \begin{align}
  \label{eq:opt_stop_sg}
    \T &= \min\{ t\in\bN:~ \cC_t \leq \alpha \}, \\
    \label{eq:opt_dec_sg}
    \sd_\T &= \arg \max_j ~(a_j-b_j\Delta_\T^j) ~ p_j\left(\{\vy_t\}_{t=1}^\T\right), \\
    \label{eq:opt_est_sg}
    \hat{\x}_\T^j &= \left(\frac{\mU_{t,j}}{\sigma^2}+\mSi_j^{-1}\right)^{-1} \left(\frac{\vv_{t,j}}{\sigma^2}+\mSi_j^{-1}\vmu_j\right),
  \end{align}
  where $\mU_{t,j}=t \mH_j'\mH_j$ and $\vv_{t,j} = \mH_j' \sum_{s=1}^t \vy_s$,
    \be
\label{eq:opt_cost_sg}
    \cC_t = \sum_{j=0}^J (a_j-b_j\Delta_t^j)\Pro_j(\sd_t\not=j) + b_j\Delta_t^j
\ee
  is the optimal cost at time $t$; 
  \be
  \label{eq:post_cost_sg}
    \Delta_\T^j=\Tr\left(\left(\frac{\mU_{\T,j}}{\sigma^2}+\mSi_j^{-1}\right)^{-1}\right)
  \ee
  is the MMSE under $\Hyp_j$ at time $\T$; 
  \be
  \label{eq:like_sg}
  p_j\left(\{\vy_t\}_{t=1}^\T\right) = \frac{\exp\left[-\frac{1}{2} \left( \sum_{t=1}^\T \frac{\|\vy_t\|^2}{\sigma^2} + \|\vmu_j\|^2_{\mSi_j^{-1}} - \big\|\frac{\vv_{\T,j}}{\sigma^2}+\mSi_j^{-1}\vmu_j\big\|^2_{\left(\frac{\mU_{\T,j}}{\sigma^2}+\mSi_j^{-1}\right)^{-1}} \right)\right]} {(2\pi)^{m\T/2}~\sigma^{m\T}~|\mSi_j|^{1/2}~\left|\frac{\mU_{\T,j}}{\sigma^2}+\mSi_j^{-1}\right|^{1/2}}
  \ee
  is the likelihood under $\Hyp_j$ at time $\T$.
\end{pro}

\begin{IEEEproof}
Since separated detection and estimation costs (cf. Section \ref{sec:sep}) are used in the problem formulation [cf. \eqref{eq:cost_sg}], from Corollary \ref{cor:sep}, when $\Hyp_j$ is decided, the optimum estimator under $\Hyp_j$ is used. For the LQG model assumed in \eqref{eq:meas_sg} and \eqref{eq:cost_sg}, the optimum estimator is the MMSE estimator, and, from \eqref{eq:cond_mean_lqg}, written as in \eqref{eq:opt_est_sg}.
 
In the previous sections, the optimum decision functions are all for binary hypothesis testing. Next we will derive the optimum decision function for the multi-hypothesis case here. 
Substituting the optimum estimator $\hat{\x}_\T^j$ in \eqref{eq:cost_sg} the estimation cost can be written as
\begin{align}
\label{eq:det_sg1}
\Exp_j\left[\|\hat{\x}_T^j-\vx\|^2 \ind{d_T=j}\right] &= \Exp_j\left[ \Exp_j\left[\|\hat{\x}_T^j-\vx\|^2|\cF_T\right] \ind{d_T=j}\right] \\
&= \Exp_j\left[ \Delta_T^j \ind{d_T=j}\right], \nn
\end{align}
where $\Delta_T^j=\Tr(\Cov_j[\vx|\cF_T])$ is the MMSE under $\Hyp_j$ at time $T$, and we used that $\ind{d_T=j}$ is $\cF_T$-measurable to write \eqref{eq:det_sg1}. From \eqref{eq:gaus_dist}, $\Delta_T^j$ is given by \eqref{eq:post_cost_sg}. Using $\Pro_j(d_T\not=j)=\Exp_j[\ind{d_T\not=j}]$ and $\ind{d_T=j}=1-\ind{d_T\not=j}$, from \eqref{eq:cost_sg}, we write the cost function as 
\be
\label{eq:cost_sg1}
    \sC(T,d_T) = \sum_{j=0}^J \Exp_j\left[ (a_j-b_j\Delta_T^j) \ind{d_T\not=j} \right] + b_j\Delta_T^j.
\ee
The optimum detector $\sd_T$ minimizes the sum of costs $\Exp_j\left[ (a_j-b_j\Delta_T^j) \ind{d_T\not=j} \right]$ when $\Hyp_j$ is rejected, i.e., 
\be
    \sd_T = \arg\min_{d_T} \sum_{j=0}^J \int\cdots\int_{\cR_j(d_T)} (a_j-b_j\Delta_T^j) ~ p_j\left(\{\vy_t\}_{t=1}^T\right) ~\text{d}\vy_1 \ldots ~\text{d}\vy_T,
\ee
where $\cR_j(d_T)\in\bR^{M\times T}$ is the subspace where $\Hyp_j$ is rejected, i.e., $d_T\not=j$, under $\Hyp_j$. To minimize the summation over all hypotheses, the optimum detector, for each observation set $\{\vy_t\}$, omits the largest $(a_j-b_j\Delta_T^j) ~ p_j\left(\{\vy_t\}_{t=1}^T\right)$ in the integral calculation by choosing the corresponding hypothesis. That is, the optimum detector is given by
\be
\label{eq:det_sg}
    \sd_T = \arg \max_j ~(a_j-b_j\Delta_T^j) ~ p_j\left(\{\vy_t\}_{t=1}^T\right),
\ee
where, from \eqref{eq:gaus1}, $p_j\left(\{\vy_t\}_{t=1}^T\right)$ is given by \eqref{eq:like_sg}.

Finally, since $\Delta_t^j$ is deterministic, from \eqref{eq:cost_sg1} and \eqref{eq:det_sg}, the optimal cost $\cC_t=\sC(\sd_t)$ is given by \eqref{eq:opt_cost_sg}.
\end{IEEEproof}

The optimal cost $\cC_t$ can be numerically computed offline for each $t$ by estimating the sufficient statistics $\{\Pro_j(\sd_t\not=j)\}_j$ through Monte Carlo simulations. Specifically, under $\Hyp_j$, we can independently generate the samples of $\vx$ and $\{\vw_1,\ldots,\vw_t\}$, and compute $\sd_t$ as in \eqref{eq:opt_dec_sg}. Then, the ratio of the number of instances where $\sd_t\not=j$ to the total number of instances gives an estimate of the probability $\Pro_j(\sd_t\not=j)$. Once the sequence $\{\cC_t\}$ is obtained, the optimum detection and estimation time is found offline using \eqref{eq:opt_stop_sg}. 

As in Section \ref{sec:sep}, the optimum detector in \eqref{eq:opt_dec_sg} is biased towards the hypothesis with best estimation performance (i.e., smallest MMSE), hence is an ML \& MMSE detector.

\subsection{Numerical Results}

\begin{figure}
  \centering
  \includegraphics[width=0.5\textwidth]{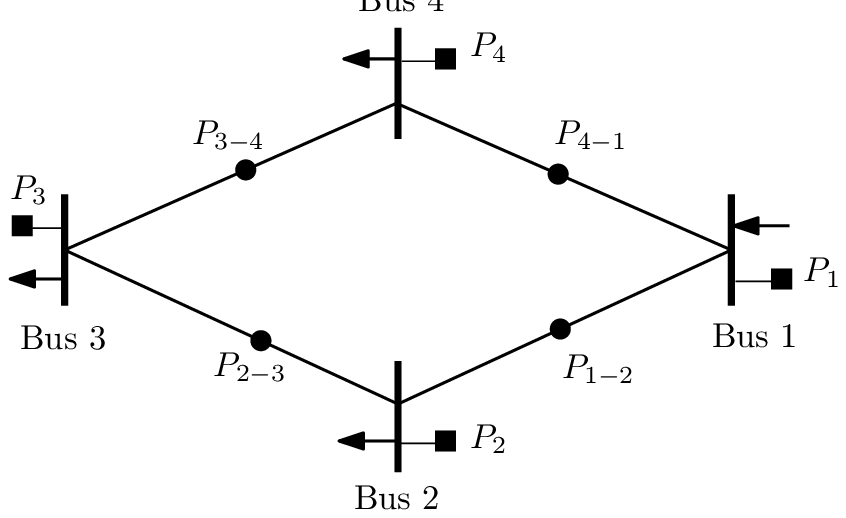}
  \caption{Illustration for the IEEE-4 bus system with the power injection (square) and power flow (circle) measurements.}\label{fig:IEEE_4bus}
\end{figure}

We next present numerical results for the proposed scheme using the IEEE-4 bus system (c.f. Fig. \ref{fig:IEEE_4bus}). Note that in this case the state status is characterized by a $3$-dimensional vector, i.e., ${\bf x}\in \mathbb{R}^3$ (the phase angle of bus $1$ is taken as the reference). In Fig. \ref{fig:IEEE_4bus}, it is seen that there are eight measurements collected by meters, thus the topology is characterized by a $8$-by-$3$ matrix, i.e., $\mH\in \mathbb{R}^{8\times 3}$.

Since the impedances of all links are known beforehand, we assume that they are of unit values without loss of generality. Here, instead of considering all possible forms of $\mH$, we narrow down the candidate grid topologies to the outage scenarios. In particular, as given in \eqref{eq:top_unc}, $\mH_0$ represents the default topology matrix, and $\{\mH_i, i=1, 2, 3, 4\}$ correspond to the scenarios where the links $\{l_{1-2}, l_{2-3}, l_{3-4}, l_{4-1}\}$ ($l_{i-j}$ denotes the link between bus $i$ and bus $j$) break down, respectively. 

We use the following distributions for the state vector $\vx$ under the hypotheses $\{\Hyp_i\}$. 
\begin{align}
\Hyp_0:&~\vx \sim \cN(\pi/5 \times \vone, \pi^2/9 \times \mI), ~~ 
\Hyp_1:~\vx \sim \cN(2\pi/5 \times \vone, \pi^2/16 \times \mI), ~~ \nn\\ 
\Hyp_2:&~\vx \sim \cN(3\pi/5 \times \vone, \pi^2/25 \times \mI), ~~ 
\Hyp_3:~\vx \sim \cN(4\pi/5 \times \vone, \pi^2/36 \times \mI), ~~ \nn\\ 
\Hyp_4:&~\vx \sim \cN(\pi \times \vone, \pi^2/4 \times \mI), \nn
\end{align} 
where $a_i=0.2, \; b_i=0.8, \forall i$, $\vone$ is the vector of ones and $\mI$ is the identity matrix. The measurements are contaminated by the white Gaussian noise $\vw_t\sim\cN(\vec{0},\mI)$. The goal is to decide among the five candidate grid topologies, and meanwhile, to estimate the state vector.
\begin{align}
&{\mH}_0=\bordermatrix{
\phantom{abc}&{\theta_2}&{\theta_3}& {\theta_4}\cr
P_{1}&-1&0&-1\cr
P_{1-2}&-1&0&0\cr
P_{2}&2&-1&0\cr
P_{2-3}&1&-1&0\cr
P_{3}&\;\;-1&2&-1\cr
P_{3-4}&0&1&-1\cr
P_{4}&0&-1&2\cr
P_{4-1}&0&0&1
},\quad
\mH_1=\left(\begin{array}{ccc}
   0& 0 & -1\\ 0 &0 &0 \\1 & -1 & 0\\ 1 &-1 &0\\-1 &2 &-1\\0 & 1 & -1\\0 &-1 &2\\0 & 0 &1
  \end{array}\right), \nn\\ \nn\\
&  \mH_2=\left(\begin{array}{ccc}
  -1 & 0 & -1\\-1 &0 &0 \\1 & 0 & 0\\ 0 & 0 & 0\\0 &1 &-1\\0 & 1 & -1\\0 &-1 &2\\0 & 0 &1
  \end{array}\right),\quad
  \mH_3=\left(\begin{array}{ccc}
  -1 & 0 & -1\\-1 &0 &0 \\2 & -1 & 0\\ 1 &-1 &0\\-1 & 1 &0\\0 & 0 & 0\\0 & 0 &1\\0 & 0 &1
  \end{array}\right), \quad
  \mH_4=\left(\begin{array}{ccc}
  -1 & 0 & 0\\-1 &0 &0 \\2 & -1 & 0\\ 1 &-1 &0\\-1 &2 &-1\\0 & 1 & -1\\0 &-1 &1\\0 & 0 & 0
  \end{array}\right).  \label{eq:top_unc}
\end{align}
\\

\begin{figure}
  \centering
  \includegraphics[width=0.9\textwidth]{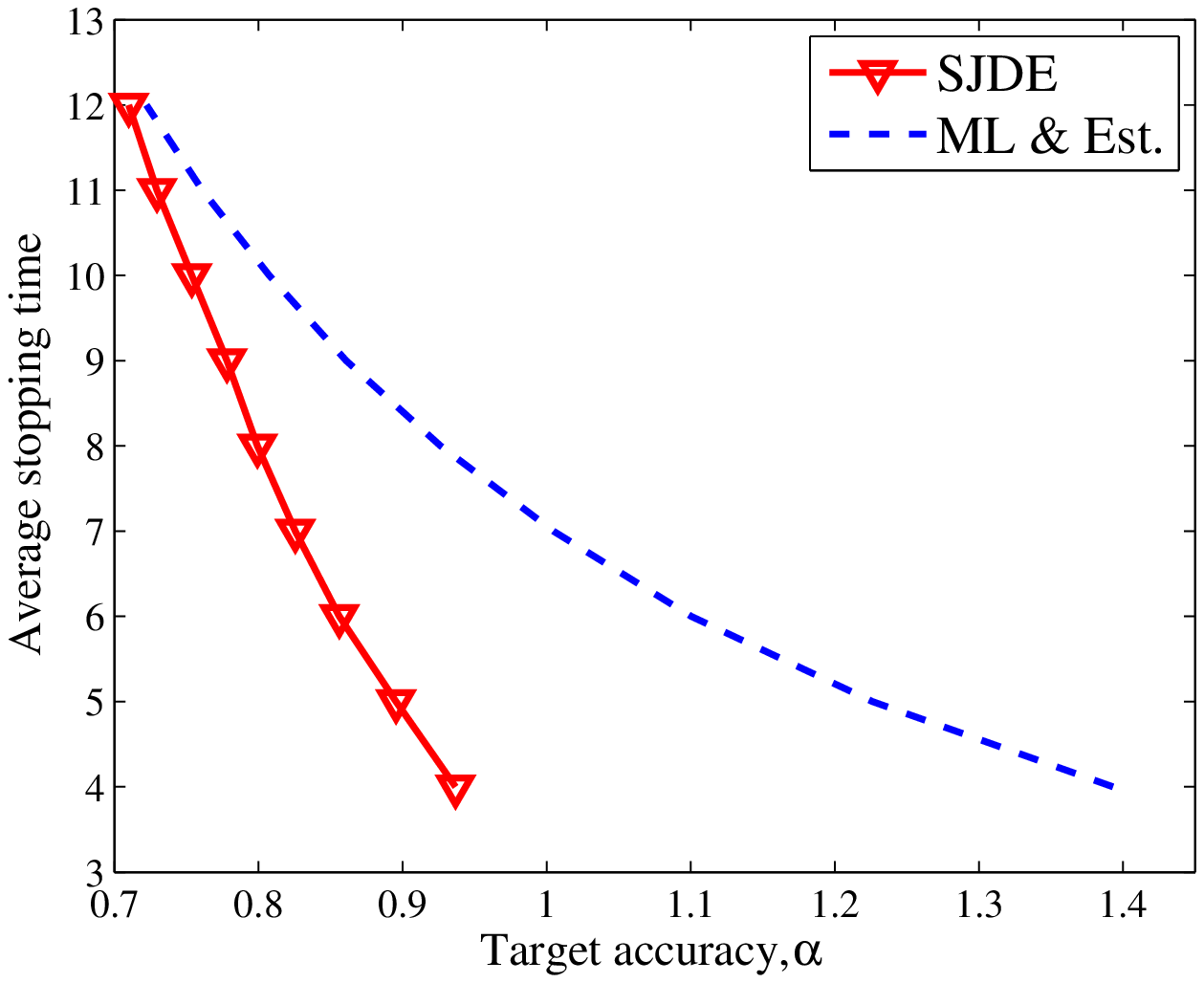}
  \caption{Average stopping time vs. target accuracy level for SJDE and the combination of ML detector \& MMSE estimator equipped with the stopping rule of SJDE.}\label{fig:fig2}
\end{figure}

Since SPRT is not applicable in the multi-hypothesis case, we compare the proposed sequential joint detection and estimation (SJDE) scheme with the combination of maximum likelihood (ML) detector and MMSE estimator, equipped the stopping time given in \eqref{eq:opt_stop_sg}. The ML detector uses the decision function
\be
\label{eq:ML_det}
d_\T = \arg \max_j ~a_j ~ p_j\left(\{\vy_t\}_{t=1}^\T\right) 
\ee
at the optimum stopping time presented in Proposition \ref{pro:sg}, hence is not a completely conventional scheme. 
Fig. \ref{fig:fig2} illustrates that SJDE [i.e., the hybrid ML \& MMSE detector, given by \eqref{eq:opt_dec_sg}] significantly outperforms this combination [i.e., the conventional ML detector in \eqref{eq:ML_det}] in terms of the overall detection and estimation performance measured by the combined cost function, introduced in \eqref{eq:cost_sg}.  
We see that SJDE requires smaller average number of samples than ML \& Est. to achieve the same target accuracy. Specifically, with small average sample size (i.e., stopping time), the improvement of SJDE is substantial. This is because smaller sample size causes larger estimation cost $\Delta_\T^j$, which in turn emphasizes the advantage of the proposed detector over the conventional ML detector. In fact, in smart grid monitoring, the typical sample size is small since the system state evolves quickly, and thus there is limited time to estimate the current state.

\section{Conclusion}
\label{sec:conc}

We have developed a general framework for optimum sequential joint detection and estimation, considering the problems in which simultaneous detection and estimation with minimal sample size is of interest. The proposed framework guarantees the best overall detection and estimation performance under a Bayesian setup while minimizing the sample size. The conventional separate treatment of the two subproblems has been shown to be strictly suboptimal since the optimum detector and estimators are strongly coupled with each other. We have also showed how the theoretical results, that are derived for a general model, apply to commonly used LQG models, including dynamic spectrum access in cognitive radio and state estimation in smart grid. We have supported the theoretical findings with numerical results.

\end{document}